\documentstyle[preprint,aps,epsf,prd,floats,amssymb]{revtex}

\def\mc{{\cal M}_{\rm cl}}
\def\mq{{\cal M}_{\Lambda}}
\def\CM{{\cal M}}
\def\tr{{\rm Tr\, }}
\def\c{{\Bbb C}}

\def\dd{\hbox{\kern0.3em/\kern-0.7em /\kern0.5em}}
\def\lie#1{{\rm Lie}\left( #1 \right)}
\def\gr{G_r}
\def\fib#1{ \pi^{-1}\left( \pi \left( #1 \right) \right) }

\def\ssqr#1#2{{\vbox{\hrule height #2pt
\hbox{\vrule width #2pt height#1pt \kern#1pt\vrule width #2pt}
\hrule height #2pt}\kern- #2pt}}

\def\sqr{\mathchoice\ssqr8{.4}\ssqr8{.4}\ssqr{5}{.3}\ssqr{4}{.3}}

\def\aa{\vbox{\hbox{$\sqr$}
\nointerlineskip\kern-.3pt\hbox{$\sqr$}}}

\begin{document}
\tightenlines

\preprint{\vbox{
\hbox{UCSD/PTH 97--14}
\hbox{hep-th/9710024}
}}
\title{Anomaly Matching Conditions and the Moduli Space of Supersymmetric
Gauge Theories}
\author{Gustavo Dotti and Aneesh V.~Manohar}
\address{Department of Physics, University of California at San Diego,\\
9500 Gilman Drive, La Jolla, CA 92093-0319}
\date{October 1997}
\maketitle
\begin{abstract}
The structure of the moduli space of $N=1$ supersymmetric gauge theories is
analyzed from an algebraic geometric viewpoint. The connection between the
fundamental fields of the ultraviolet theory, and the gauge invariant composite
fields of the infrared theory is explained in detail. The results are then used
to prove an anomaly matching theorem. The theorem is used to study anomaly
matching for supersymmetric QCD, and can explain all the known anomaly matching
results for this case.
\end{abstract}
\pacs{PACS}

\section{Introduction}

One important constraint on the moduli space of vacua of supersymmetric gauge
theories~\cite{seiberg,susy} is that the massless fermions in the low energy
theory should have the same flavor anomalies as the fundamental fields, i.e.\
the 't~Hooft consistency conditions should be satisfied~\cite{thooft}. These
conditions are considered a particularly stringent test on the spectrum of
massless fermions, which is usually obtained from symmetry arguments and
renormalization group flows. It is found that for some theories the classical
moduli space $\mc$ or a suitable quantum modified version $\mq$ of it satisfies
't~Hooft consistency conditions at every point. Other theories fail 't~Hooft's
test at some vacua, and it is believed that in the infrared, these theories 
correspond to the weak coupling sector of a dual theory with a different gauge
group and matter content~\cite{susy}. In this paper, sufficient conditions on
the fundamental chiral field content $\phi^i$ for a theory to satisfy the
't~Hooft conditions is established. This is usually done by first finding a
basic set of gauge invariants $\hat \phi ^j (\phi)$ and the constraints among
them, then checking point by point the matching of flavor anomalies between the
fundamental fields and the composites at each point on the moduli space. This
procedure involves tedious calculations, and does not offer a systematic
approach to the problem of determining whether $\mc$ (or a suitable quantum
modified version of it) gives the right description of the infrared sector or
not. Our results establish simple sufficient conditions on the fundamental
fields that guarantee 't~Hooft consistency conditions will be satisfied. The
novelty of our approach is that no explicit calculation of anomalies is
required at any time; it is not even required to know what the basic gauge
invariants are.

This paper fills in the details omitted in Ref.~\cite{dm}. The outline of the
paper is as follows. We devote Sec.~\ref{sec:vacua} to a review of the
connection between $\mc$ and the algebraic quotient of $U$ under the action of
the complexification $G$ of the gauge group. The results in~\cite{lt} are
rederived and additional information related to the structure of gauge orbits
in $U$ and the geometry of $\mc$ is provided. Methods to determine the
dimension of $\mc$ before finding the invariants and their constraints are
presented. A number of examples illustrating how naive expectations fail to be
true in special situations is also given in this section. In Sec.~\ref{sec:uv},
we analyze anomaly matching at points on the classical moduli space. The proof
of the anomaly matching theorem makes use of the connection between $\mc$ and
the vector space $U$ spanned by the fundamental matter fields $\phi^i$ provided
by the map $\pi: U \to \mc$, $\pi(\phi) = \hat \phi^i(\phi)$. Knowledge of the
algebraic geometric construction of $\mc$ and details related to this map is
essential. In Sec.~\ref{sec:ir} we give a rigorous proof of the fact that the
flavor anomalies of the massless modes at the vacuum $\hat \phi_0$ equal those
of the full vector space $V$ of the gauge invariant composites (i.e.\ ignoring
all constraints) when ${\cal M} $ is the set of critical points of an invariant
superpotential. We apply this result to: (i) extend the matching theorem to
cases where the superpotential $W$ is not trivial, (ii) globalize the
point-by-point result of the matching theorem, and (iii) prove that anomaly
matching conditions are compatible with integrating out fields.

Our results guarantee the matching of flavor anomalies between the UV and $\mc$
for the large family of $s$-confining theories introduced in~\cite{sconf} and
those theories obtained from them by integrating out matter fields, which have
a quantum modified moduli space. As an application, in Sec.~\ref{sec:examples}
we analyze in detail the well known case of supersymmetric QCD. We first repeat
the analysis of Ref.~\cite{seiberg}, and find all points at which anomalies
match by performing explicit calculations, then show how the matching follows
readily from our results with virtually no calculations.  Extensions of the
results to dual theories will be described elsewhere.

\section{Supersymmetric Vacua and Algebraic Geometry}\label{sec:vacua}

In this section, we will review some of the properties of supersymmetric vacua
in supersymmetric gauge theories, and their connection with algebraic geometry.
Many of the results are well known in the physics or mathematics literature.
Including the results here will allow us to define our notation, and also to
make the paper more self-contained.

\subsection{SUSY Gauge Theories}

The physical objects that we will consider in this paper are supersymmetric 
gauge theories with gauge group $\gr$, where $\gr$ is the direct product of a
compact connected semisimple Lie group and (possible) $U(1)$ factors. We will
assume that all the $U(1)$'s are compact, i.e. that the fields have rational
charges. The action of the gauge theory is
\begin{equation}\label{2.1}
S[\phi] = \int d^4x\, d^4 \theta\ \phi^\dagger e^V \phi +
 \int d^4x\, d^2\theta\ \left[{1\over 4 g^2} \tr W^\alpha W_{\alpha} 
+ W(\phi) + c.c. \right]
\end{equation}
where $\phi$ is a set of chiral superfields transforming as a (reducible)
representation of $\gr$, $W_{\alpha}$ is the gauge chiral superfield, $g$ is
the gauge coupling constant, $V$ is the gauge vector superfield, 
\begin{equation}
V = V^A T^A,
\end{equation}
$T^A$ are the generators of the Lie algebra of $G_r$, and $W$ is the
superpotential. $V$ and $W_\alpha$ are related by $W_\alpha = -(1/4) \bar D
\bar D D_\alpha V$. For simplicity of notation, we have assumed that $\gr$ is a
simple Lie group with coupling constant $g$; if not, there is a different
coupling constant for each simple factor in $G_r$. The action Eq.~(\ref{2.1})
will in general also have a global flavor symmetry group $F$, and a $U(1)_R$
symmetry.

The set of inequivalent vacua of a supersymmetric gauge theory is referred to 
as the moduli space. The classical moduli space $\mc$ is determined by studying
gauge inequivalent constant field configurations that are critical points of
the superpotential $W$, and satisfy the $D$-flatness condition
\begin{equation}\label{dflat}
\phi^\dagger T^A \phi = 0\ \ \forall\ \ T^A \in \lie{\gr},
\end{equation}
where $\lie{\gr}$ is the Lie algebra of $\gr$, and has dimension $d_G$. The
action Eq.~(\ref{2.1}) has a larger invariance than the gauge symmetry $\gr$.
It is also invariant under local transformations of the form
\begin{equation}\label{gi}
\phi \to e^{i \Lambda^A T^A} \phi, \qquad\qquad
e^{C^A T_A} \to e^{i (\Lambda^A)^{\dagger} T^A} e^{C^A T^A}
e^{-i \Lambda^A T^A},
\end{equation} 
with $\Lambda^A$ chiral superfields~\cite{wb}. This implies invariance under
$G$, the complexification of $\gr$. This result was used in Ref.~\cite{lt} to
show that the moduli space of supersymmetric vacua is an algebraic variety. We
will review the analysis given in Ref.~\cite{lt} here because we need
additional details about the construction of the moduli space and the structure
of $G$-orbits not discussed previously.

Any classical supersymmetric vacuum configuration can have only a constant 
expectation value for the scalar component of the superfield. Thus classical
supersymmetric vacua are a subset of $U$, the vector space of all constant
field configurations $\phi$. $U$ has dimension $d_U$, the number of chiral
superfields. If $\phi$ is a point in $U$, the $G$-orbit of $\phi$ will be 
denoted by $G \phi$. If $\phi$ is a critical point of the superpotential $W$,
then so are all points of $G \phi$ since $W$ is $G$-invariant. The set of
critical points of $W$ in $U$ will be denoted by $U^W$.

\subsection{Algebraic Geometry}

The mathematical objects that we will consider in this paper are affine
algebraic sets (a special case of varieties) over the field of complex 
numbers.\footnote{We will use the notation of Refs.~\cite{shaf,gw}.} Let
$\c\left[x_1,\ldots,x_n\right]$ be the ring of polynomials in $n$ complex
variables $x_1,\ldots,x_n$. Let $p_\alpha\left(x_1,\ldots,x_n\right)$,
$\alpha=1,\ldots,k$ be a finite set of polynomials in the $n$ variables. Then
the algebraic set $V\left(p_\alpha\right)$ is the set $\left\{x_i \in \c\, | \,
p_\alpha(x)=0\ \forall\ \alpha \right\}$. It can be thought of as a curve in
$\c^n$ given implicitly by the polynomial equations $p_\alpha=0$. The ideal
$(p_1,\ldots,p_k)$ is the ideal generated by the polynomials $p_\alpha$, i.e.
the set of polynomials of the form $\sum_\alpha f_\alpha\, p_\alpha$, where
$f_\alpha$ are arbitrary polynomials in $\c\left[x_1,\ldots,x_n\right]$.
Clearly, any polynomial in $(p_1,\ldots,p_k)$ vanishes at all points on
$V(p_\alpha)$. The set of polynomials that vanish on an algebraic set $X$ (such
as $V(p_\alpha)$) will be denoted by $I(X)$, and forms a finitely generated
ideal of $\c\left[x_1,\ldots,x_n\right]$. In general, $I(V(p_\alpha)) \supseteq
(p_1,\ldots,p_k)$, but equality need not hold. It is possible to define the
algebraic set $X$ as the zero set of polynomials $g_i$, $i=1,\ldots,s$ that
generate $I(X)$. In this case, $X=V(g_i)$, and $I(X)=(g_1,\ldots,g_s)$. If this
is the case we say that the equations $g_i(x_1,...,x_n) = 0$, $i=1,...,s$ {\em
correctly define} $X$. 

An irreducible algebraic set is one that cannot be written as a proper union of
two algebraic sets. Any algebraic set can be written as a finite union of 
irreducible algebraic sets in a unique way. The tangent space $T_pX$ at a point
$p$ of the algebraic set $X \subseteq {\Bbb C}^n$ correctly defined by
polynomials $g_i(x)$ is the vector subspace $\ker ( \partial g_i / \partial x_j
)_p$ of ${\Bbb C}^n$. The dimension $d_X$ of an irreducible algebraic set $X$
is $d_X = \min_{p \in X} \dim \; T_p X$. There are alternative equivalent
definitions of $T_pX$ and different ways of calculating dimensions of
irreducible algebraic sets (see \cite{gw,shaf}). The natural complex valued
functions on an algebraic set $X$ are the restrictions to $X$ of polynomials in
its ambient vector space. These are called regular functions, and the set of
regular functions is the coordinate ring $\c[X]$. Note that two polynomials
$f_1$ and $f_2$ in $\c[x_1,...,x_n]$ define the same regular function on $X$ if
and only if $f_1 - f_2 \in I(X)$. This defines an equivalence relation in
$\c[x_1,...,x_n]$ and, clearly, $\c[X] = \c[x_1,...,x_n]/I(X).$ Regular
functions from the algebraic set $X \subseteq {\Bbb C}^n$ to the algebraic set
$Y \subseteq{\Bbb C}^m$ are naturally defined to be those that can be written
as the restriction to $X$ of $m$ polynomials in ${\Bbb C}^n$. All the geometric
properties of $X$ are encoded in $\c[X]$, and $X$ can be constructed from
$\c[X]$. This fact plays a key role in the discussion of the next subsection. 

It is natural to use the Zariski topology in studying algebraic geometry. The
Zariski-closed sets of $\c^n$ are algebraic sets, by definition. The fact that
this correctly defines a topology in ${\Bbb C}^n$ is non trivial, and is a 
consequence of the Hilbert basis theorem, which implies that an infinite 
intersection of algebraic sets is an algebraic set. In the rest of
this paper, we will mainly use the Zariski topology, so open and closed sets
will always mean with respect to the Zariski topology. In a few places, we will
also need to use the more familiar metric-topology on $\c^n$, and we will state
this explicitly. The Zariski topology might seem a little strange to readers
used to thinking about the more familiar metric topology. In $\c$, for example,
the Zariski-closed sets are $\c$, a finite set of points, or the null set. Thus
the Zariski-closure of the set of integers in $\c$ is the entire space $\c$. In
contrast, the metric-closure of the set of integers is itself. Regular
functions $f: X \to Y$ from the algebraic set $X \subseteq \c^n$ into the
algebraic set $Y \subseteq \c^m$ are continuous in the induced Zariski
topology, whereas most other functions which are continuous in the metric
topology fail to be Zariski continuous. Thus, the Zariski topology allows us to
get the strongest results when dealing with regular function. As an example, a
regular function $f: \c \to \c$ that vanishes on the integers ${\Bbb Z}$, must
vanish on the closure $\overline{{\Bbb Z}} = \c$, which  is a fancy way of
saying that the only complex polynomial with infinite roots is the trivial one. 

The natural groups to study in algebraic geometry are linear algebraic groups.
Let $M_n(\c)$ be the space of all $n\times n$ complex matrices.
$GL(n,\c)$ is the set $\det M \not=0$ in $M_n(\c)$. $G \subseteq GL(n,\c)$ is a
linear algebraic group if it is the intersection of $GL(n,\c)$ with an
algebraic set in $M_n(\c)$. The group $SU(n)$ is not an algebraic group,
because the constraint that the matrix be unitary is not a polynomial relation;
it involves complex conjugation. The complexification of $SU(n)$ is the group
$SL(n,\c)$, which is an algebraic group since the condition that the
determinant be unity is a polynomial relation in the elements of the matrix.
The complexification of all compact connected semisimple Lie groups are
algebraic groups. The vacuum structure of supersymmetric gauge theories will
depend on $G$ , the complexification of the gauge group $\gr$. $G$ is an
algebraic group, but $\gr$ need not be.

\subsection{SUSY QED}\label{sec:sqed}

\begin{figure}
\epsfxsize=8cm
\centerline{\epsffile{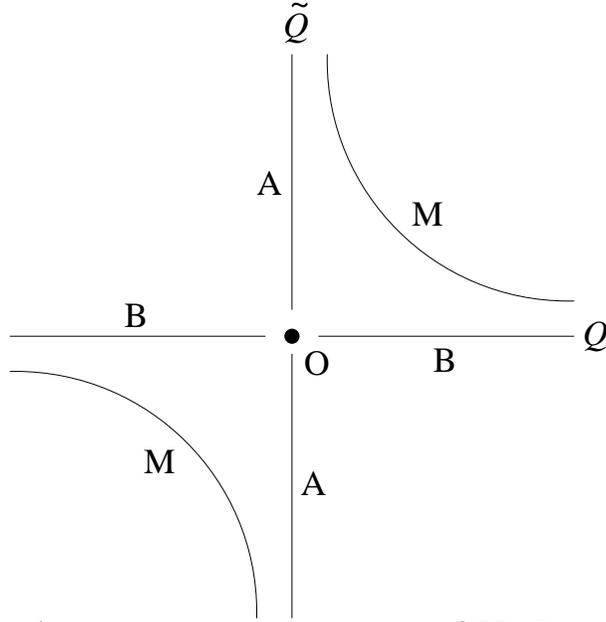}}
\caption{The structure of gauge orbits in supersymmetric QED. The orbits are
connected. They appear disconnected because the figure only shows the
restriction of the configuration space to real values of $Q$ and $\tilde Q$.
The orbits $M$ and $O$ are closed, and $A$ and $B$ are not closed.  $M$, and $A
\cup B \cup O$ are fibers.
\label{fig:SQED}}
\end{figure}

It is instructive at this point to go over the well-known example of SUSY QED,
which is a $U(1)$ supersymmetric gauge theory with two chiral superfields $Q$
and $\tilde Q$ with charges $\pm 1$, respectively. The set of all constant
field configurations is $\c^2=\left\{ \left( Q, \tilde Q \right) \right\}$ for
SQED. In SQED, the group $G=U(1)^c$ is the set of transformations $(Q,\tilde Q)
\rightarrow (z Q, \tilde Q/z)$, where $z\in \c$, $z\not=0$, as can be seen from
Eq.~(\ref{gi}). The $G$ orbits are illustrated in Fig.~\ref{fig:SQED}. They are
the curves $M = \left\{(Q,\tilde Q) \ |\ Q\tilde Q=M\not=0 \right\}$, the curve
$A=\left\{(Q,0)\ |\ Q\not =0 \right\}$, the curve $B=\left\{(0,\tilde Q)\ | \
\tilde Q\not=0\right\}$, and the origin $O=\left\{(0,0)\right\}$. The
decomposition of $U$ into $G$ orbits is quite complicated, even for the simple
example of SQED. $G$ orbits need not be closed. The orbits $M$ and $O$ are
closed, and the orbits $A$ and $B$ are not. The closure of the $G$ orbit of
$\phi$ is denoted by $\overline {G\phi}$. A very useful result is that the
closure of a $G$ orbit in the Zariski topology is equal to its closure in the
metric topology. The closure of the two orbits $A$ and $B$ are
the curves $(Q,0)$, and $(0,\tilde Q)$, where $Q$ and $\tilde Q$ are no longer
restricted to be non-zero. The closure $\overline{G \phi}$  of orbits $G \phi$
of an algebraic group are irreducible algebraic sets. The closures of
$G$-orbits are given by the polynomial equations $Q \tilde Q =M$ for $\bar M$,
$\tilde Q=0$ for $\bar A$, $Q=0$ for $\bar B$, and $Q=0, \tilde Q=0$ for $\bar
O$, each of which is an irreducible algebraic set.

\subsection{Algebraic Quotients and the Moduli Space}

The power of algebraic geometry lies in the interplay between geometric
properties of the algebraic set $X$ and algebraic properties of its coordinate
ring $\c[X]$. The key properties of the ring $\c[X]$ are: (i) $\c[X]$ has no
nilpotents (that is, $f \in \c[X]$, $f^m = 0$ implies $f = 0$), and (ii)
$\c[X]$ is finitely generated. It turns out (\cite{shaf}, chap.~I, sect 2, Thm
1) that these are also sufficient conditions for a given ring ${\cal A}$ over
$\c$ to be isomorphic to the coordinate ring of some algebraic set $X_{{\cal
A}}$, ${\cal A} \cong \c[X_{{\cal A}}]$. The construction of $X_{{\cal A}}$ is
in fact very simple: take a set $t_1,...,t_n$ of generators of ${\cal A}$. In
general, ${\cal A}$ is not a free algebra; there are polynomial equations among
the generators $g_i(t_1,...,t_n) = 0 , i=1,...,k$. One shows that $X_{{\cal A}}
= V(g_i) \subseteq \c^n$, the equality being valid only in the case of a free
algebra. Now assume $X = U^W \subseteq U$, the $G$ invariant algebraic set of
critical points of a polynomial $G$ invariant superpotential $W$, and let $d_U$
be the dimension of $U$. There is a natural representation of $G$ on $\c[U^W]$,
namely $\c[U^W] \ni f \to g \circ f, g \circ f(x) \equiv  f(g^{-1}x), g \in G$.
Under this representation the homogeneous polynomials of degree $d$ form
invariant subspaces. For reductive linear algebraic groups ($G$ is reductive if
any regular representation is completely reducible), the subring $\c[U^W]^G$ of
$G$ invariant polynomials is finitely generated, as follows from Thm.~4.1.1 in
\cite{gw}. As $\c[U^W]^G$ does not have nilpotent elements, it can be thought
of as the coordinate ring of an algebraic set $U^W \dd G = X_{\c[U^W]^G}$. If
$p_1^G,...,p_n^G$ is a minimal set of generators of $\c[U^W]^G$, then $U^W \dd
G \subseteq \c^n$ is the algebraic set defined by the constraints between the
$p_i^G$'s. There is a natural regular map $\pi: U^W \to U^W \dd G$ by $U^W \ni
(x_1,...,x_{d_U}) \to (p_1^G(x_1,...,x_{d_U}),...p_n^G(x_1,...,x_{d_U}))$
obtained by evaluation of the generating polynomials. The pair $(\pi, U^W \dd
G)$ is called algebraic (or categorical) quotient of $U^W$ under the action of
$G$ in the mathematics literature. The basic property of the quotient is that,
by construction,  $\pi_* : \c[U^W \dd G] \to \c[U^W]^G$ is an isomorphism (as
algebras over ${\Bbb C}$) between the coordinate ring of $U^W \dd G$ and the
ring of invariant polynomials in $U^W$. Therefore, given any algebraic variety
$Z$ and $G$-invariant regular map $f: U^W \to Z$, there exists a unique regular
map $\tilde{f}: U^W \dd G \to Z$ such that $f = \tilde{f} \circ \pi$, a
property that uniquely determines the quotient. Further details of this
construction, including properties one through three quoted without proof
below, can be found in problems 11.2.6-1,2 in \cite{gw}, and also in \cite{IV}.
Using these, we prove other results required to understand the
ultraviolet-infrared global anomaly matching. We assume $G$ is reductive, as is
always the case in  physically interesting theories. Property P5 establishes
the anticipated contact with supersymmetric moduli spaces: points in $U^W \dd G
$ are in one to one correspondence with closed orbits in $U^W$. As the latter
are in one to one correspondence with physically inequivalent classical
supersymmetric vacua, $U^W \dd G$ is the classical moduli space $\mc$ of the
theory (\ref{2.1}). This picture breaks if $G$ fails to be reductive
(subsection~\ref{subsecF}). The fiber of $\pi\left( \phi \right)$,
$\pi^{-1}\left(\pi\left(\phi\right)\right)$ is in general a reducible algebraic
set. It can be written as the union $\pi^{-1}\left(\pi\left(\phi\right)\right)
= X_1\cup \ldots \cup X_n$ of $G$-invariant irreducible algebraic sets. (Note
that $X_i$ are $G$-invariant, but need not be a single $G$-orbit.) Then $X_1
\cap \ldots \cap X_n$ is a closed $G$ invariant subset of
$\pi^{-1}\left(\pi\left(\phi\right)\right)$, and contains the unique closed $G$
orbit.

One can now connect the algebraic construction of $U^W \dd G$ discussed above
with the classical moduli space $\mc$ of supersymmetric vacua~\cite{lt}. In
Ref.~\cite{lt} it is proven that every $G$ orbit contains a (unique) $D$-flat
point if and only if the orbit is closed. Every fiber contains exactly one
closed $G$ orbit, and thus exactly one $D$-flat point which is also a critical
point of $W$. Thus the gauge-inequivalent field configurations that satisfy the
$D$-flatness condition  and $\nabla W=0$ are in one-to-one correspondence with
fibers $\pi^{-1}\left(\pi\left(\phi\right) \right)$. This means that $U^W \dd
G^c$ is isomorphic to $\mc$ the classical moduli space of the supersymmetric
gauge theory, the result shown in Ref.~\cite{lt}.

In studying anomaly matching conditions, one needs to compute anomalies in the
full theory (UV theory) and for the massless modes (IR theory). The UV
anomalies are computed using the fields $\phi$, i.e.\ using $U$, and the IR
anomalies are computed using the classical moduli space $\mc$ or its quantum
modification $\mq$. The classical moduli space $\mc \sim U \dd G$, so the
anomaly matching conditions require knowledge of the structure of the
$G$-orbits in $U$.

\subsection{Collection of Mathematical Results}

We will collect here a number of useful results from Refs.~\cite{gw,IV} that
will be needed later in this article. Some of them have already been discussed
earlier in this section. The basic setting is the algebraic quotient $\pi: U^W
\rightarrow U^W \dd G$ of the variety $U^W$ given by the critical points of $W$
under the action of a reductive algebraic group $G$.

\begin{description}

\item[(P1)] $\pi$ is surjective

\item[(P2)] If $C \subset U^W$ is closed and $G$-invariant, then $\pi(C)$ is
closed in $U^W \dd G$ and $\left(\pi|_C,\pi\left(C\right)\right)$ is the 
algebraic quotient of $C$ under the action of $G$.

\item[(P3)] If $\left\{Z_i\right\}_{i\in I}$ is any collection of closed
$G$-invariant subsets of $U^W$, then $\pi\left(\cap_i Z_i\right) = \cap_i 
\pi\left(Z_i\right)$.

\item[(P4)] If $U^W$ is irreducible, $U^W \dd G$ is irreducible. The proof
follows from (P1) and Lemma A.1.16 of Ref.~\cite{gw}.

\item[(P5)] There is a unique closed $G$ orbit in every fiber of the map $\pi$.

\item[(P6)] Two points $\phi_i, i=1,2$ lie in the same fiber if and only if
$\overline{G \phi_1} \cap \overline{G \phi_2} \neq \emptyset$.
\footnote{Fibers are called {\em extended orbits} in \cite{lt}. This 
characterization of fibers follows from {\bf (P3)} and the fact that 
$\pi(\overline{G \phi})$ consists of a single point.} 

\end{description}

\noindent From now on, we shall assume that $U^W$ is irreducible. Then $\mc = 
U^W \dd
G$ is irreducible, and we can use the results of Chap.~1 Sec.~6.3 
of \cite{shaf}.

The dimensions of various objects are defined as follows:

\begin{description}

\item[$d_G$:] The real dimension of the Lie Algebra $\lie {\gr}$ 
of the gauge group $\gr$, which is equal to the complex dimension of $\lie G$.

\item[$d_U$:] The dimension of $U$. 

\item[$d_{U^W}$:] The dimension of $U^W$.

\item[$d_V$:] The number of generators of the ring of $G$-invariant polynomials
$\c[U^W]^G$.

\item[$d_{\cal M}$:] The dimension of the classical moduli space $\mc = U^W \dd
G \subseteq V = \c^{d_V}$.

\item[$d$:] The maximum dimension of a $G$-orbit in $U^W$. (Note that
$G$-orbits can have different dimensions, so $d$ is the maximum possible
dimension. Also, $d \le d_G$.) 

\item[$d_f$:] The minimum dimension of a fiber in $U^W$.

\end{description}

Let $T^A, A=1,\ldots,d_G$ be a basis for $\lie{\gr}$, the Lie Algebra of the
gauge group $\gr$. The dimension of a $G$ orbit through $\phi$ equals the rank
of the $d_U \times d_G$ matrix $A(\phi)$ with columns $T^A \phi$. Note that the
entries of $A(\phi)$ are polynomials of degree one in $\phi$. Let ${\cal
A}_j(\phi)$, $j \leq \min(d_G,d_U)$ be the set of all $j \times j$ minors of
$A(\phi)$, and ${\cal I}_j$ the set of polynomials in $\phi$ obtained by taking
the determinants of the elements of ${\cal A}_j$. The rank of $A(\phi)$ equals
the maximum value of $j$ for which there is a polynomial in ${\cal I}_j$ which
does not vanish at $\phi$. The set $O_{<j}$ of points in $U$ whose orbits have
dimension less than $j$ is the closed set $O_{<j} = V({\cal I}_j)$, obtained by
requiring that all $j \times j$ minors have zero determinant. The complement is
then an open set. Hence ``maximum rank'' of a matrix valued function on an
algebraic set with regular entries is then an example of an {\em open
condition,} i.e.\ it defines an open set. The same type of argument shows that
the set of smooth points of an irreducible algebraic set $Y$ is open in $Y$.
These results together with a straightforward application of~\cite{shaf}
Chap.~1, Sec~6.3 imply

\begin{description}

\item[(P7)] The following is a list of non-empty open subsets of $U^W$ and 
$\mc$. Quantities with a $\, \hat{} \,$ belong to $\mc$, and without a $\,
\hat{} \,$ belong to $U^W$. The inclusion sign means the left-hand side is a
non-empty open set contained in the right-hand side, but equality need not
hold.
\begin{description}
\item[] $\hat O_1 \subseteq
\left\{ \hat \phi \in \mc \, | \, \dim \left( 
\pi^{-1}(\hat \phi ) \right) = d_{U^W} - d_{\cal M} \right\}.$
\item[] $\hat O_2 = \left\{ \hat \phi \in \mc \, | \, 
\hat \phi \ {\rm is\ a\ regular\ point\ of}\ \mc \right\}$.
\item[] $\hat O_3 = \left\{ \hat \phi \in \mc \, | \, \dim \left( 
\pi^{-1} (\hat \phi ) \right) = d_f \right\}$.
\item[] $O_i=\pi^{-1}(\hat O_i),\ i=1,2,3$.
\item[] $O_4 = \left\{ \phi \in U \, | \dim \left( G\phi \right) = d \right\}$.
\end{description}
\end{description}
Using these non-empty open sets, and the result that non-empty open subsets of
an algebraic set are dense, and any two non-empty open sets have a non-empty
intersection (Lemma A.1.12 in \cite{gw}) allows one to prove 
\begin{description}
\item[(P8)] The dimension of $\mc$ equals that of $U^W$ minus the minimum
dimension of a fiber. Proof: take a point $\hat \phi \in \hat O_1 \cap \hat
O_3$. Then $d_{U^W} - d_{\cal M}=d_f$.
\item[(P9)] The minimum dimension of fibers is greater than or equal to the
maximum dimension of $G$-orbits. Proof: Take $\phi \in O_3 \cap O_4$. As $\pi$
is $G$-invariant, $T_\phi G\phi \subseteq T_\phi \fib{\phi}$, so that $d \le
d_f$. (Here $T_p X$ denotes the tangent space of $X$ at the point $p$.)
\end{description}
We finally quote an important theorem due to Knop~\cite{knop}, and prove a
result that is used in Sec.~\ref{sec:examples}

\noindent{\bf Theorem 1 [Knop]} (Ref.~\cite{knop})\ \  If $G$ is semisimple,
$G\phi$ has maximum dimension, and $\pi\left(\phi\right)$ is smooth, then
$\pi'_\phi $ is surjective.

\medskip

\noindent{\bf Theorem 2} If $\dim\left(G\phi_0\right)=d$ (maximal)
 and $G\phi_0$ is closed, then
\begin{itemize}
\item[(a)] $\pi^{-1}(\pi(\phi_0)) = G\phi_0.$
\item[(b)] $d_{\cal M} = d_{U^W} - d$.
\end{itemize}
 If also $d = d_G$ then
\begin{itemize}
\item[(c)] $\pi'_{\phi_0}$ is surjective.
\end{itemize}
Finally, if $\phi_0$ 
is also a smooth point
\begin{itemize}
\item[(d)] $\ker \pi'_{\phi_0} = \lie G \phi_0$ and $\pi(\phi_0)$ is smooth. 
\end{itemize}

\smallskip
\noindent Proof: Assume (a) is false, and pick any point $\phi_1 \in
\pi^{-1}(\pi(\phi_0)) \setminus G \phi_0$. Because of $\bf (P6)$,
$\overline{G\phi_1} \cap G\phi_0 \not= \emptyset$. Therefore there is a
sequence $g_n \in G$ and a point $g_0 \in G$ such that $g_n \phi_1 \rightarrow
g_0 \phi_0$. Given any point $g \phi_0 \in G\phi_0$, $g g_0^{-1} g_n \phi_1
\rightarrow g \phi_0$, so $g \phi_0 \in \overline{G\phi_1}$. Thus $G\phi_0$  is
a proper subset of $\overline{G\phi_1}$, which implies that $\dim \left(
G\phi_0 \right) < \dim \left( G\phi_1 \right)$ (Lemma A.1.18 of \cite{gw}), 
i.e. $\dim \left( G\phi_1 \right) > d$ which is a contradiction, since
$G$-orbits have maximum possible dimension $d$.

(b) $\bf (P9) $ and (a) implies $d_f = d$, and (b) follows using $\bf (P8)$.

(c) By Luna's theorem \cite{luna} $G\phi_0$ closed implies there is an \'etale
slice at $\phi_0$. In the case dim~$G \phi_0 = d_G$ the existence of the slice
implies that any curve $\hat \phi (t)$ through $\hat \phi _0 = \hat \phi (0)$
can be lifted to a curve $\phi (t)$ in $U^W$ satisfying $\phi(0) = \phi_0$ and
$\pi(\phi (t)) = \hat \phi (t)$. This is equivalent to the surjectivity of
$\pi'_{\phi_0}$.

(d) If $\phi_0$ is smooth then dim $T_{\phi_0}U^W = d_{U^W}$. From (b) and (c)
we have: $d_{\cal M} \leq \text{dim} \; T_{\pi(\phi_0)}\mc = \text{rank} \;
\pi'_{\phi_0} = d_{U^W} - \text{dim ker}\; \pi'_{\phi_0} \leq d_{U^W} -
\text{dim} \; T_{\phi_0} G \phi_0 = d_{U^W} - d = d_{\cal M}.$ Thus, these are
all equalities and (d) follows.

\noindent
Conditions (c) and (d) ensure that the tangent to the moduli space is
isomorphic to the massless modes of the theory determined from the ultraviolet
Lagrangian at the classical level, i.e. by looking at the massless modes of the
quadratic part of the Lagrangian in unitary gauge.

\subsection{Examples}\label{subsecF}

\subsubsection{SUSY QED}

We have already discussed this case. The space $X$ is $\c^2$, since there are
two complex fields $Q$ and $\tilde Q$. The ring $\c[X]$ is the ring of
polynomials in two indeterminates, $\c[Q,\tilde Q]$. The ring $\c[X]^{G}$ is
the set of gauge invariant polynomials $\c[Q\tilde Q]$, freely generated by the
single polynomial $M=Q \tilde Q$. The algebraic set $Y$ defined by
$\c[Y]=\c[Q\tilde Q] = \c[X]^G$ is thus $\c$. The map $\pi:X\rightarrow Y$
takes $(Q,\tilde Q)$ to the point $M=Q\tilde Q$ in $Y$. The fiber of a point
$M\not=0 \in Y$ is the closed orbit $M$ discussed in Sec.~\ref{sec:sqed}. The
fiber of $M=0$ is $A \cup B \cup O$. Note that $\c=Y=X \dd U(1)^c = \c^2 \dd
U(1)^c$ is a smooth algebraic set, with no singular points. Nevertheless the
fiber of $M=0$ is different in structure from that of the points $M\not=0$. The
fibers $\pi^{-1}(M)$ are closed irreducible and smooth one-dimensional
algebraic sets for $M\not=0$. $\pi^{-1}(0)$ is the reducible and singular
algebraic set $Q\tilde Q=0$, and contains three $G$ orbits, $A$, $B$ and $O$.
It can be written as the union of two irreducible components, $X_1$ given by
$Q=0$ and $X_2$ given by $\tilde Q=0$. Their intersection, $X_1 \cap X_2$ is
the origin $O=(0,0)$, and is the unique closed $G$ orbit in $\pi^{-1}(0)$. The
smooth points $Z$ of $\pi^{-1}(0)$ form the open set $A \cup B$. The complement
$\pi^{-1}(0) \setminus Z$ is the origin $O$ and is a smooth closed algebraic
set that contains the unique closed $G$ orbit. $\pi'$ is surjective at all
points of $\pi^{-1}(M \neq 0)$ and at all points of the fiber $\pi^{-1}(0)$
except $(0,0)$. Thus it is possible in this case to choose a point in
$\pi^{-1}(0)$ where $\pi'$ is surjective; however this point is not in the
unique closed orbit $(0,0)$ in the fiber.

Now consider adding a superpotential $W=Q \tilde Q$ to SQED. The critical
points of $W$ are $Q=\tilde Q =0$, so that $U^W$ is a single point, and is
$G$-invariant. Note that in this example, the critical points of $W$ {\sl do
not} form a complete fiber. The ideal $I(V(\nabla W))$ is the set of all
polynomials of the form $Q f(Q,\tilde Q) + \tilde Q g (Q, \tilde Q)$, and
$I(V(\nabla W))^G$ is the set of all polynomials of the form $Q \tilde Q h(Q
\tilde Q)$, where $f$, $g$ and $h$ are arbitrary polynomials in their
arguments. The classical moduli space $\mc$ without a superpotential is the
complex plane $\c$ given by $M = Q \tilde Q$. Including $W$ restricts one to
$\mc^W \subseteq \mc$ given by the algebraic set $M=0$ in $\mc$. Since $W$ is
gauge invariant, one can rewrite $W$ in terms of the gauge invariant
polynomials used to describe the moduli space. In the present example, $W=Q
\tilde Q = M$. However, when regarded as a function of the gauge invariants
rather than the fundamental fields, the critical points of the resultant $W$ do
not correctly describe the moduli space in the presence of a superpotential. In
our example, $W=M$ has no critical points, whereas $\mc^W$ contains the single
point $M=0 \in \mc$. One can obtain $\mc^W$ by minimizing $W=M^2$, instead of
$W=M$. However, in general, one cannot obtain the equations defining the moduli
space $\mc^W$ by minimizing a superpotential in the space of gauge invariants.

\subsubsection{SUSY QED with equal charges}

\begin{figure}
\epsfxsize=8cm
\centerline{\epsffile{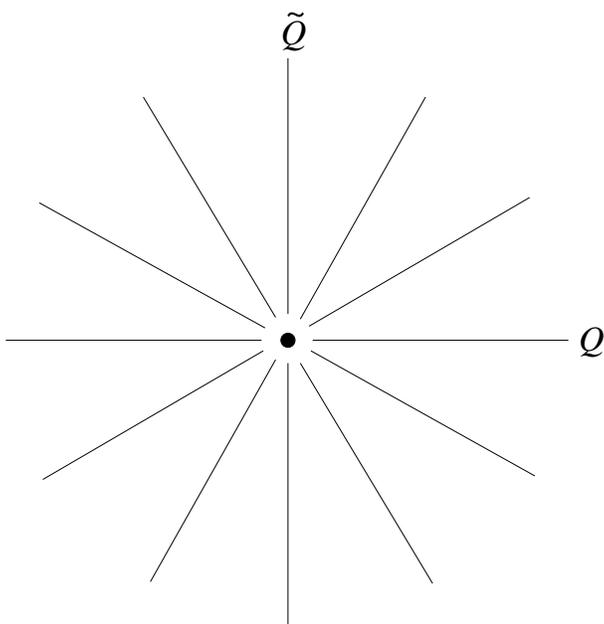}}
\caption{The structure of gauge orbits in supersymmetric QED with equal
charges. Only the projection onto real values of $Q_1$ and $Q_2$ is shown. The
orbits of a generic point $(Q_1,Q_2)$ are radial lines, and are not closed. The
orbit of $(0,0)$ is a single point, and is closed. There is a single fiber
which is $\c^2$.
\label{fig:SQEDp}}
\end{figure}

An interesting example is SUSY QED with fields $Q_1$ and $Q_2$ both with charge
$+1$. This is an anomalous theory, but still provides a useful example that
illustrates the structure of fibers. Similar results can be found in more
complicated anomaly free theories (such as SUSY QCD with $N_f=N_c$). The space
$U$ is $\c^2$, and $\c[U]=\c[Q_1,Q_2]$. There are no gauge invariant
polynomials other than constants, so $\c[U]^G = \c$, and the moduli space $\mc
=U \dd U(1)^c$ is the zero dimensional space consisting of a single point $P$.
The orbits of all points other than $(0,0)$ are radial lines (see
Fig.~\ref{fig:SQEDp}), and are not closed.  The orbit of $(0,0)$ is a single
point, and is the unique closed orbit. The fiber $\pi^{-1}(P)$ is the entire
plane $\c^2$, and is irreducible. In this case $d_U=2$, $d_V=0$, $d_{\cal M} =
0$, $d_G=1$, $d=1$ and $d_f=2$. Note that $d_f > d$, so that the minimum
dimension of a fiber can be strictly greater than the maximum dimension of an
orbit.

\subsubsection{An example where $\pi'$ is not surjective for an orbit of
maximal dimension}

This example shows that Theorems~1 and 2 are the most we can say about
surjectivity of $\pi'$ above smooth points of $\mc$. Consider a $U(1)$ theory
with three fields $Q_2$, $Q_+$ and $Q_-$ with charges 2, 1 and $-1$ (an
anomalous theory). The space $U$ is $\c^3$. The gauge invariant polynomials are
generated by $A=Q_+ Q_-$ and $B=Q_2 Q_-^2$. The moduli space $\mc$ and $V$ are
both $\c^2$, since there are no relations among $A$ and $B$. $\mc$ is smooth
everywhere. Consider the orbit of $\phi = (0,Q_+,0)$. It is the set of all
points $(0,z Q_+,0)$ with $z\not=0$, and has dimension one, i.e. equal to the
dimension of the gauge group. However,
\begin{equation}
{\rm rank}\ \pi' = {\rm rank}
\left( \begin{array}{ccc}
0 & Q_- & Q_+ \\
Q_-^2 & 0 & 2 Q_2 Q_- \\
\end{array} \right) \le 1\ ({\rm if}\ Q_- = 0),
\end{equation}
so $\pi'$ is not surjective even though the orbit has maximal dimension and
lies above a smooth point. Note that Theorem~1 does not apply here because the
gauge group is not semisimple, and Theorem~2 does not apply here, because the
orbit is not closed.

\subsubsection{SUSY QCD} 

This is the example studied in detail in Refs.~\cite{seiberg,susy}. The gauge
group is $\gr = SU(N)$, with $N_F$ matter fields $Q^{i \alpha}$,
$i=1,\ldots,N_F$, $\alpha=1,\ldots,N$ in the fundamental $N$ representation of
$SU(N)$, and $N_F$ matter fields $\tilde Q_{\beta j}$, $j=1,\ldots,N_F$,
$\beta=1,\ldots,N$ in the $\bar N$ representation of $SU(N)$. The ultraviolet
space $U$ is a vector space of dimension $d_{U} = 2 N N_F$. The
complexification of $SU(N)$ is $G=SL(N,\c)$, under which $Q^{i \alpha}$ and
$\tilde Q_{j \beta}$ are in the fundamental and its dual
representation.\footnote{Given a representation of a group $H$ in a vector
space $V$ by $v^i \to h^i_j v^j$ we define the dual and conjugate
representations on $V^*$ by $w_i \to w_j (h^{-1})^j_i$ and $w_i \to w_j
(h^{\dagger})^j_i$. They agree only when the representation of $H$ on $V$ is
unitary. Note that the dual representation is defined so as to make $w_i v^i$
invariant, which is not the case for the conjugate of a non-unitary
representation.} A set of generators of $\c[U]^G$, the coordinate ring of gauge
invariant polynomials are the mesons and baryons,
\begin{eqnarray} \label{QCD} 
M^i_j &=& Q^{i \alpha} \tilde Q_{\alpha j} \nonumber \\ 
B_{k_1 \cdots k_{(N_F-N)}} &=& {1\over N!} Q^{i_1 \alpha_1} Q^{i_2 \alpha_2} 
\cdots Q^{i_N \alpha_N} \epsilon_{\alpha_1 \alpha_2
\cdots \alpha_N} \epsilon_{i_1 i_2 \cdots i_N k_1 \cdots k_{(N_F-N)}} \\
\tilde B^{l_1 \cdots l^(N_F-N)} &=& {1\over N!} 
\tilde Q_{\alpha_1 j_1} \tilde Q_{\alpha_2 j_2} \cdots
\tilde Q_{\alpha_N j_N} \epsilon^{\alpha_1 \alpha_2 \cdots \alpha_N} 
\epsilon^{j_1 j_2 \cdots j_N l_1 \cdots l_{(N_F-N)}}. \nonumber 
\end{eqnarray}
These polynomials span the vector space $V$. The structure of the classical
moduli space $\mc \subseteq V$ depends crucially on the value of $N_F$.

\begin{description}

\item[{\bf (i)} $\mathbf{N}_{\mathbf{F}}\, {\bf < }\, \mathbf{N}$:] If $N_F <
N$, $B$ and $\tilde B$ are identically zero. $\c[U]^G$ is freely generated by
$M^i_j$, $\mc = V$ and $d_{{\cal M}} = N_F^2$. This example illustrates a
non-trivial case of dimension counting. Consider the point $\phi$ given by 
\[
Q^{i \alpha} = \tilde Q_{\alpha i} = \left\{
\begin{array}{ll}
\delta^{i \alpha} & i \leq N,\\
0 & \mbox{otherwise.}
\end{array}\right.
\]
The $SL(N,\c)$ orbit of $\phi$ is closed and of maximum dimension, $\dim \;
G\phi = \dim \, SL(N,\c) - \dim \, SL(N-N_F,\c) = 2NN_F - N_F^2$. From
Theorem~2 we obtain $\dim \, \mc = d_U - \dim \, G\phi = N_F^2$. 
\item[{\bf (ii)}$\mathbf{N}_{\mathbf{F}}\, {\bf = }\, \mathbf{N}$:]
In this case $B = \det(Q), \tilde B = \det(\tilde Q)$, and $\c[U]^G$ is not a
free algebra, as its generators are constrained by the single relation
\begin{equation} \label{nf=n}
\det M - B \tilde B = 0.
\end{equation}
This gives a hypersurface in $V \cong \c^{N^2 + 2}$, and the dimension of the
moduli space is $d_{{\cal M}} = N^2 +1$. This number can also be obtained by
applying theorem~2 to the closed $G$ orbit of the point $(Q^{i \alpha} =
\delta^{i \alpha}, \tilde Q = 0)$

\item[{\bf (iii)}$\mathbf{N}_{\mathbf{F}}\, {\bf = }\, \mathbf{N} $ {\bf + 1 
with a superpotential}:] When $N_F=N+1$ the fields (\ref{QCD}) are subject to
the following algebraic constraints
\begin{equation}
\label{CMSN+1}
\text{Cof}(M)^j_i - \tilde B^j B_i = 0, \hspace{.4cm}
M^i_j B_i = 0, \hspace{.4cm} M^i_j \tilde B^j = 0,
\end{equation}
where cof$(M)^i_j$ is the matrix of cofactors.  Assume we add a superpotential
$W =m Q^{N+1 \alpha} \tilde Q_{\alpha N+1}$, which is a mass term for the
$N+1^{\rm th}$ flavor. The set $U^W$ of critical points of $W$ is just the
space $U$ for $N_F=N$, naturally embedded in $U$ for $N_F=N+1$ by setting the
components of $Q$ and $\tilde Q$ for the $N+1^{\rm th}$ flavor to zero.
$\pi(U^W)$ is the intersection of the moduli space Eq.~(\ref{CMSN+1}) with the
subspace $B_i = \tilde B^i = M^{N+1}_i = M^i_{N+1} = 0, i=1,...N$ This
reproduces Eq.~(\ref{nf=n}), as anticipated by P2 above.
\end{description}

\subsubsection{An example involving a non-reductive group}

\begin{figure}
\epsfxsize=8cm
\centerline{\epsffile{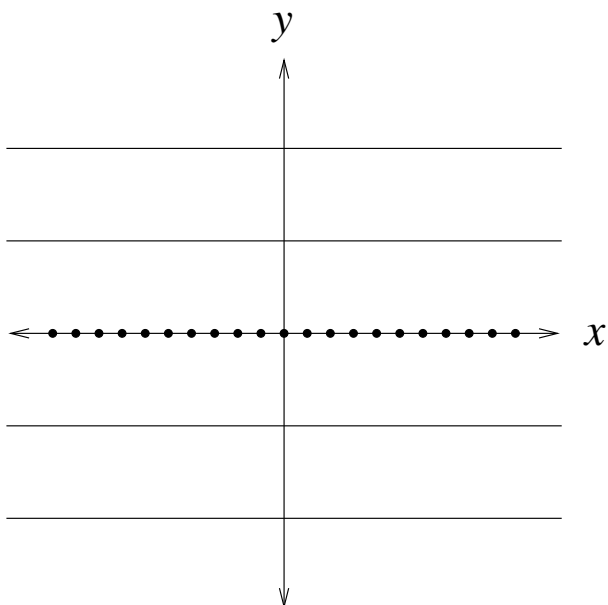}}
\caption{The structure of gauge orbits for the non-reductive group $\c^+$.  The
orbits are horizontal lines if $y \neq 0$, and points if $y=0$. The fibers are
the horizontal lines $y={\rm constant}$. The fiber $y=0$ contains an infinite
number of closed orbits.
\label{fig:nonred}}
\end{figure}

This example is from~\cite{IV}. The abelian group $ G = \c^+$ of complex
numbers under addition (which is the complexification of the group $\gr$ of
real numbers under addition) is a simple example of a non-reductive group. We
will consider the representation on $U = \c^2$ given by $(x,y) \to (x + zy, y),
z \in \c^+$. Note that $(x,0)$ is an invariant subspace with no invariant
complement. $\c^+$ is the linear algebraic non-reductive subgroup of $GL(2,\c)$
of upper triangular, determinant one matrices of the form 
\begin{equation} 
\left( \begin{array}{cc}
1 & z \\
0 & 1 \\
\end{array} \right), 
\end{equation}
acting on $\c^2$ as the restriction of the fundamental representation of
$GL(2,\c)$. The orbits $G (x_0,y_0)$ are closed one dimensional lines $(x,y_0)$
when $y_0 \neq 0$. For $y=0$ the orbits are points $(x_0,0)$ for each value of
$x_0$ (see Fig.~\ref{fig:nonred}). Every $G$ orbit is closed. The fibers are
the horizontal lines $y=y_0$. The $x$-axis $y=0$ is a fiber which contains an
infinite number of closed orbits. If this example were an acceptable
supersymmetric gauge theory, the algebraic quotient $U \dd G$ would not equal
the classical moduli space $\mc$, since the fiber $y=0$ contains infinitely
many closed orbits, i.e. infinitely many inequivalent supersymmetric vacua.

\section{Anomaly Matching between the Ultraviolet and Infrared Theories}
\label{sec:uv}

\noindent{{\bf Theorem 3:}}\ \ Let $\mc$ be the classical moduli space of a
supersymmetric gauge theory with gauge group $G_r$ and flavor symmetry $F$ and
superpotential $W$. It is assumed that the gauge theory has no gauge or
gravitational anomalies, and the flavor symmetries have no gauge anomalies. Let
$\hat \phi_0 \in \mc$ be a point in the classical moduli space. Assume there is
a point $\phi_0 \in U^W$ in the fiber $\pi^{-1}(\pi(\phi_0))$ of $\hat \phi_0$
such that
\begin{itemize}
\item[(a)] $G$ (the complexification of $G_r$) is completely broken at
$\phi_0$, so that $\lie G \phi_0 \cong \lie G$.
\item[(b)] $\ker \pi^\prime_{\phi_0} = \lie G \phi_0$ and
$\pi^\prime_{\phi_0}$ is surjective.
\end{itemize}
If a subgroup $H \subseteq F$ is unbroken at $\hat \phi_0$, then the 't~Hooft
consistency conditions for the $H^3$ flavor anomalies and the $H$ gravitational
anomalies are satisfied.

For the purposes of the proof, it is convenient to write the original flavor
symmetry as $F' \times R$, where $R$ is the $R$-symmetry, and $F'$ now contains
only non-$R$ symmetries. We first prove anomaly matching when $H \subseteq F'$
and $W=0$, and then prove consistency for anomalies that include the $R$
symmetry. (Note that the unbroken $R$ symmetry might be a linear combination of
the original $R$ symmetry and some generator in $F'$.) The extension to
$W\not=0$ follows simply from the results of Sec.~\ref{sec:ir}.

Since $H$ is unbroken at $\hat \phi_0$, $\hat \phi_0$ is $H$-invariant
\begin{equation}
\lie H \hat \phi_0 =0.
\end{equation}
The map $\pi : U \rightarrow \mc$ commutes with the flavor symmetries, so
\begin{equation}
0=\lie H \hat \phi_0 = \lie H \left( \pi \left( \phi_0 \right) \right)
=\pi^\prime_{\phi_0}\left( \lie H \phi_0 \right).
\end{equation}
Thus, by (a) and (b)
\begin{equation}
\lie H \phi_0 \subseteq \ker \pi^\prime_{\phi_0} \cong \lie G .
\end{equation}
This implies that given any ${\frak h} \in \lie H$, there is a unique ${\frak g
\left( h \right )} \in \lie G$ such that 
\begin{equation}\label{3.1}
{\frak h} \phi_0 = - {\frak g \left( h \right )} \phi_0,
\end{equation}
where the minus sign is chosen for convenience. It is straightforward to check
that the map $\lie H \rightarrow \lie G$ given by $\frak h \rightarrow \frak g
\left( h \right )$ is a Lie-algebra homomorphism,
\begin{equation}
\frak g \left( \left[ h_1, h_2 \right] \right ) = \left[ g \left( h_1 \right ),
g \left( h_2 \right ) \right].
\end{equation}
This allows us to define a new ``star'' representation of $\lie H$ in $U$
\begin{equation}
{\frak h}^* \equiv {\frak h} + {\frak g \left( h \right )}.
\end{equation}
Since $\lie G \phi_0 \in \ker \pi_{\phi_0}^\prime$, the new $\lie H$
representation on $T_{\hat \phi_0} \mc$ defined by $\pi'_{\phi_0}{\frak h}^*$
agrees with the original one. Thus the $\frak h^*$-anomalies computed at $\hat
\phi_0 \in \mc$ are the same as the $\frak h$-anomalies at the same point.

$\lie G \phi_0$ is an invariant subspace under ${\frak h}^*$, and the
restriction of ${\frak h}^*$ to $\lie G \phi_0$ is the adjoint action by
$\frak g \left( h \right )$. This can be seen by direct computation. Take any
element ${\frak g} \phi_0 \in \lie G \phi_0$. Then
\begin{equation}
{\frak h}^* {\frak g} \phi_0 = {\frak h g} \phi_0 + {\frak g\left( h \right) g}
\phi_0 = {\frak g h} \phi_0 + {\frak g\left( h \right) g} \phi_0 = 
{\left[ \frak g \left( h \right), g \right] } \phi_0 = 
Ad_{\frak g \left( h \right )}\, {\frak g} \phi_0,
\end{equation}
since the flavor and gauge symmetries commute, and using Eq.~(\ref{3.1}). The
space $U$ can be broken up into the tangent space to the $G$-orbit $T_{\phi_0}
G\phi_0 = \lie G \phi_0 \cong \lie G$ and its invariant complement,
$C_{\phi_0}$, since $G$ is reductive. By (b), the map $\pi^\prime_{\phi_0}$ is
a bijective linear map from $C_{\phi_0}$ to the tangent space $T_{\hat \phi_0}
\mc$ of the moduli space $\mc$ at $\hat \phi_0$, and commutes with $\frak h^*$.
Thus the action of $\frak h^*$ on $C_{\phi_0}$ is equivalent to the action of
$\frak h$ on $T_{\hat \phi_0} \mc$, by the similarity transformation $S$ given
by $\pi^\prime_{\phi_0}$ restricted to $C_{\phi_0}$. One can write
\begin{equation}\label{3.a}
\frak h^* = {\frak h}_{\rm UV} + {\frak g \left( h \right )} =
\left( \begin{array}{cc}
S\, {\frak h}_{\rm IR}\, S^{-1} & 0 \\
0 & Ad_{\frak g \left( h \right )} \\
\end{array} \right),
\end{equation}
where the second form shows the structure of $\frak h^*$ on $U = C_{\phi_0}
\oplus T_{\phi_0} G\phi_0$. The action of $\frak h$ on $U$ has been labeled by
the subscript UV, and the action on the moduli space has been labeled by IR.

One can now compare anomalies in the UV and IR theories using the two different
forms for $\frak h^*$. Since the adjoint representation is real, the
$\left(\frak h^*\right)^3$ flavor anomaly and $\frak h^*$ gravitational anomaly
are equal to the anomalies in the infrared theory. All that remains is the
proof that the $\left( \frak h^* \right)^3$ and $\frak h^*$ anomalies of $U$
equal the ${\frak h}^3$ and $\frak h$ anomalies of $U$. Let $\frak h^{\rm
A,B,C}$ be any three elements of $\lie H$. Then
\begin{eqnarray}
\tr \frak h^{* \rm A} \left\{ \frak h^{* \rm B}, \frak h^{* \rm C} \right\}
&=& {\phantom +} \tr \frak h^{\rm A}_{\rm UV} \left\{ \frak h^{\rm B}_{\rm UV}, 
\frak h^{\rm C}_{\rm UV} \right\} \nonumber \\
&& + \tr \frak g\left(h^{\rm A}\right)  \left\{ \frak h^{\rm B}_{\rm UV}, 
\frak h^{\rm C}_{\rm UV} \right\} + {\rm cyclic} \nonumber \\
&& + \tr \frak h^{\rm A}_{\rm UV} \left\{ \frak g\left(h^{\rm B}\right), 
\frak g\left( h^{\rm C} \right) \right\} + {\rm cyclic} \nonumber \\
&&+\tr \frak g\left( h^{\rm A} \right) \left\{ \frak g\left( h^{\rm B} \right), 
\frak g\left( h^{\rm C} \right) \right\}
\end{eqnarray}
The last three lines vanish because the original theory had no gauge and 
gravitational anomalies, and the flavor symmetries have no gauge anomalies.
Thus the $\frak h^3$ and $\left( \frak h^* \right)^3$ anomalies are the same.
Similarly the $\frak h^*$ and $\frak h$ anomalies agree since $\frak g$ is
traceless because there is no gravitational anomaly. Thus 't~Hooft's
consistency condition for the flavor anomalies is satisfied.

We now prove the matching theorem for anomalies involving the $R$-charge using
an argument similar to the one presented above. The $R$-charge acting on $U$ is
given by the matrix $\frak r$. The $R$-charge is defined acting on chiral
superfields, and so is the charge of the scalar component. Anomalies are
computed using the fermionic components, so it is convenient to define a new
charge $\tilde {\frak r}$ which we will call fermionic $R$-charge, defined by
\begin{equation}
\tilde {\frak r} = {\frak r} -1.
\end{equation}
The anomaly can be computed by taking traces over the chiral superfields of
$\tilde {\frak r}$. The reason for making the distinction between $\frak r$ and
$\tilde {\frak r}$ is that the map $\pi$ from $U$ to $\mc$ commutes with
$R=\exp \frak r$, but does not commute with $\tilde R = \exp \tilde{\frak r}$.

Assume that $R$ is unbroken at $\hat \phi_0 = \pi\left( \phi_0 \right)$. Then
by an argument similar to that above, it is possible to define a ``star'' 
$R$-charge, $\frak r^*$,
\begin{equation}\label{3.2}
\frak r^* \equiv r + g\left( r \right)
\end{equation}
which has the form
\begin{equation}\label{3.b}
 \frak r^* = r_{\rm UV} + g \left( r \right ) =
\left( \begin{array}{cc}
S\, {\frak r_{\rm IR}}\, S^{-1} & 0 \\
0 & Ad_{\frak g \left( r \right )} \\
\end{array} \right)
\end{equation}
under the decomposition of $U$ into $C_{\phi_0} \oplus T_{\phi_0} G \phi_0$. As
in Eq.~(\ref{3.a}), we have used the subscripts UV and IR to denote the $R$
charges in the ultraviolet and infrared theories. Note that $S$ is the same
matrix in Eqs.~(\ref{3.a},\ref{3.b}), given by $\pi^\prime_{\phi_0}$ restricted
to $C_{\phi_0}$. The fermionic $R$-charge is then given by
\begin{equation}\label{14}
 \tilde {\frak r}^* = {\frak r}^* -1 = \tilde {\frak r}_{\rm UV} + 
 {\frak g \left( r \right )} =
\left( \begin{array}{cc}
S\, \tilde {\frak r}_{\rm IR}\, S^{-1} & 0 \\
0 & Ad_{\frak g \left( r \right )} -1 \\
\end{array} \right)
\end{equation}
where in the last equality we have used the fact that fermion $R$ charge
$\tilde {\frak r}_{\rm IR} = {\frak r}_{\rm IR}-1$ in the infrared theory.

Compute the trace of $(\tilde {\frak r}^*)^3$ in $U$,
\begin{equation}
\tr \left( \tilde {\frak r}^* \right)^3 = \tr \left\{ \tilde {\frak r}_{\rm UV} 
+ {\frak g\left( r \right)} \right\}^3
= \tr \left\{ \tilde {\frak r}^3_{\rm UV} + 3 \tilde {\frak r}^2_{\rm UV} 
{\frak g\left( r \right)} + 3 \tilde {\frak r}_{\rm UV} 
{\frak g\left( r \right)}^2 + {\frak g\left( r \right)}^3 \right\}.
\end{equation}
The $R$-charge has no gauge anomaly, so $\tr_U\, \tilde {\frak r}_{\rm UV} 
\{{\frak g_{\rm A}, g_{\rm B}}\} + \tr_{\lie G}\, \{Ad_{\frak g_{\rm A}},
Ad_{\frak g_{\rm B}}\} =0$, for any ${\frak g}_{\rm A,B} \in \lie G$. Here the
first term is the matter contribution to the anomaly, and the second term is
the gaugino contribution. The absence of gauge anomalies implies that odd
powers of $\frak g\left( r \right)$ vanish when traced over the matter fields,
since there is no gaugino contribution to these anomalies. Thus we find
\begin{equation}\label{3.5}
\tr_{U} \left( \tilde {\frak r}^* \right)^3 = \tr_{U} 
\left( \tilde {\frak r}_{\rm UV} \right)^3 - 3\, \tr_{\lie G}\,
 Ad_{\frak g\left( r \right)}^2.
\end{equation}
The block diagonal form of $\tilde{\frak r}^*$ Eq.~(\ref{14}), gives
\begin{equation}\label{3.11}
\tr_U \left( \tilde {\frak r}^* \right)^3 = \tr \left( \tilde 
{\frak r}_{\rm IR} \right)^3
- \tr_{\lie G} \left( 1 + 3 Ad_{\frak g\left( r \right)}^2 \right).
\end{equation}
The $R^3$ anomaly $A_{\rm UV}\left( R^3 \right)$ in the UV theory is given by 
adding the matter and gaugino contributions
\begin{equation}\label{3.10}
A_{\rm UV}\left( R^3 \right)=
\tr_U \left( \tilde {\frak r}_{\rm UV} \right)^3 + \tr_{\lie G} 1^3 = 
\tr_U \left( \tilde {\frak r}^* \right)^3 + \tr_{\lie G} \left( 1 + 3 Ad_{
\frak g \left( r \right)}^2 \right).
\end{equation}
The $R^3$ anomaly $A_{\rm IR}\left( R^3 \right)$ in the IR theory is given by
\begin{equation}\label{3.12} A_{\rm IR}\left( R^3 \right) = \tr \left( \tilde
{\frak r}_{\rm IR} \right)^3,  \end{equation} since there are no gauginos in
the low energy theory. Combining Eq.~(\ref{3.5}--\ref{3.12}), one sees
immediately that the UV and IR anomalies are equal, $A_{\rm UV}\left( R^3
\right)=A_{\rm IR} \left( R^3 \right)$.

It is straightforward to check that the gravitational $R$ anomaly, and the $H^2
R$ and $H R^2$ anomalies match. One finds from Eqs.~(\ref{3.a},\ref{3.b}) that
\begin{equation}
\tr \tilde {\frak r}^* = \tr \tilde {\frak r}_{\rm UV} = 
\tr \tilde {\frak r}_{\rm IR} - \tr_{\lie G} 1,
\end{equation}
which is the matching condition for the gravitational $R$ anomaly, when
rewritten as $\tr \tilde {\frak r}_{\rm UV} + \tr_{\lie G} 1 = \tilde {\frak
r}_{\rm IR}$. For the $R H^2$ anomaly:
\begin{eqnarray}
\tr \tilde {\frak r}^* \left\{ {\frak h}^{*\rm A} , {\frak h}^{*\rm B} \right\}
&=& \tr \tilde {\frak r}_{\rm UV} \left\{ {\frak h}^{\rm A}_{\rm UV} , 
{\frak h}^{\rm B}_{\rm UV} \right\} 
+ \tr \frak g(r) \left\{g( h^{\rm A} ) , h^{\rm B}_{\rm UV} \right\} \nonumber\\
&&\qquad + 
\tr {\frak g(r) \left\{ h^{\rm A}_{\rm UV} , g( h^{\rm B}) \right\}}+
\tr \tilde {\frak r}_{\rm UV} \left\{ g( h^{\rm A}) , g(h^{\rm B}) \right\}
\nonumber \\
&=& \tr \tilde {\frak r}_{\rm UV} \left\{ h^{\rm A}_{\rm UV} , 
h^{\rm B}_{\rm UV} \right\} -
\tr_{\lie G} \left\{ Ad_{\frak g( h^{\rm A})} , Ad_{\frak g(h^{\rm B})} \right\}
\end{eqnarray}
where the last line follows using the fact that there are no $H G^2$ anomalies
and that the $R G^2$ anomaly cancels when the matter and gaugino contributions
are added. Using the block diagonal forms of $r^*$ and $h^*$, one has
\begin{eqnarray}
\tr \tilde {\frak r}^* \left\{ \frak h^{*\rm A} , h^{*\rm B} \right\}
&=& \tr \tilde {\frak r}_{\rm IR} \left\{ \frak h^{\rm A}_{\rm IR} , 
h^{\rm B}_{\rm IR} \right\} 
-\tr_{\lie G} \left\{ Ad_{\frak g( h^{\rm A})} , Ad_{\frak g(h^{\rm B})} 
\right\},
\end{eqnarray}
which gives the anomaly matching condition for the $R H^2$ anomaly when
combined with the previous result. Similarly, for the $R^2 H$ anomaly:
\begin{eqnarray}
\tr \left( \tilde {\frak r}^* \right)^2 \frak h^{*}
&=& \tr \tilde {\frak r}_{\rm UV}^2 {\frak h}_{\rm UV}
+ \tr \left\{ \frak g(r), g(h) \right\} \tilde {\frak r}_{\rm UV} +
\tr \frak g(r)^2 h_{\rm UV} \\
&=& \tr \tilde {\frak r}_{\rm UV}^2 {\frak h}_{\rm UV} 
-\tr_{\lie G} \left\{ Ad_{\frak g(r)}, Ad_{\frak g(h)} \right\}.
\end{eqnarray}
The block diagonal forms of $\frak r^*$ and $\frak h^*$ give
\begin{eqnarray}
\tr \left( \tilde {\frak r}^* \right)^2 {\frak h}^{*}
&=& \tr \tilde {\frak r}_{\rm IR}^2 {\frak h}_{\rm IR}
-\tr_{\lie G} \left\{ Ad_{\frak g(r)} , Ad_{\frak g(h)} \right\}.
\end{eqnarray}
Comparing with the previous equation shows that the $R^2 H$ anomalies are the
same in the UV and IR theories.

\section{The Infrared Sector}\label{sec:ir}

The results of the previous section allow one to study the matching of
anomalies between the ultraviolet and infrared theories at certain points in
the classical moduli space. In this section, we derive some results that allow
us to relate the anomaly matching conditions at different points on the moduli
space. The moduli space is no longer restricted to be the classical moduli
space $\mc$.

We consider the case where the moduli space $\CM$ is an algebraic
curve in an ambient vector space $V$ given as the critical points of a
superpotential $W$ with $R$-charge two,
\begin{equation}\label{ir:2}
\CM = \left\{ \hat \phi \in V \, | \, W_i(\hat \phi )=0 \right\},
\end{equation}
where $\hat \phi$ denotes a point in $V$, and we will use the notation
$W_i\equiv \partial W / \partial \hat \phi^i$, $W_{ij}\equiv \partial^2 W /
\partial \hat \phi^i\partial \hat \phi^j$, etc. The tangent space to $\CM$ at
$\hat \phi_0$, $T_{\hat \phi_0} \CM$, is defined by
\begin{equation}\label{ir:1}
T_{\hat \phi_0}\CM = 
\left\{ \hat v^i \in V \, | \, W_{ij}(\hat \phi )\hat v^j=0 \right\}.
\end{equation}
In all the cases we are interested in, $W$ is a polynomial in $\hat \phi$ and
Eq.~(\ref{ir:2}) correctly describes the algebraic set, so that
Eq.~(\ref{ir:1}) agrees with the algebraic geometry notion of the tangent
space.

Assume that a subgroup $H$ of the flavor symmetry group $F$ is unbroken at a
point $\hat \phi_0 \in \CM$. The invariance of the superpotential $W$ under $F$
implies that
\begin{equation}
W ( h^i_j \hat \phi_0^j ) = W (\hat \phi_0 ),
\end{equation}
where $h^i_j$ is the matrix for the $H$ transformation in the representation
$R$ of the fields $\hat \phi$. Differentiating this equation twice with respect
to $\hat \phi$ and evaluating at the $H$-invariant point $\hat \phi_0$ gives
\begin{equation}
h^k_i h^l_j W_{kl} \left( h^i_j \hat \phi_0^j \right) = 
h^k_i h^l_j W_{kl} \left( \hat \phi_0^j \right) = W_{ij}(\hat \phi_0 ),
\end{equation}
which shows that $W_{ij} ( \hat \phi_0 )$ is a $H$ invariant tensor that
transforms as $\left( \bar R \otimes \bar R \right)_S$ under $H$, where $R$ is
the $H$ representation of the low energy fields $\hat \phi \in V$. The tangent
space to $\CM$ at $\hat \phi_0$ is the null-space of $W_{ij}$. One can write $V
= T_{\hat \phi_0} \CM + N_{\hat \phi_0} \CM$ as the direct sum of the tangent
space and its orthogonal complement in $V$. Then $W_{ij}$ provides a
non-singular invertible map from $N_{\hat \phi_0} \CM$ into its dual, so that
$N_{\hat \phi_0} \CM$ transforms as a real representation of $H$. This
immediately implies that the $H$ anomalies computed using the flat directions $
T_{\hat \phi_0} \CM$ agree with those computed using the entire vector space
$V$.

A similar result holds for the anomalies involving the $R$ charge. Let $R_i$ be
the $R$-charge of $\hat \phi_i$, so that
\begin{equation}
W\left( e^{i \alpha R_i} \hat \phi_i \right) = e^{2 i \alpha} 
W \hat \phi_i )
\end{equation}
since $W$ has $R$ charge two. Differentiating twice with respect to $\hat \phi$
shows that
\begin{equation}
e^{i \alpha \left( R_i +R_j\right)}
W_{ij} (\hat \phi_0 ) = e^{2 i \alpha }
W_{ij} (\hat \phi_0 ),
\end{equation}
which can be written in the suggestive form
\begin{equation}\label{ir:3}
e^{i \alpha \left( \left[R_i-1\right] + \left[ R_j - 1 \right] \right)}
W_{ij} (\hat \phi_0 ) =
W_{ij} (\hat \phi_0 ).
\end{equation}
$R_i-1$ is the $R$ charge of the fermionic component of the chiral superfield.
Thus Eq.~(\ref{ir:3}) shows that $N_{\hat \phi_0} \CM$ transforms like a real
representation under $\tilde R = R-1$, the fermionic $R$ charge. Thus the $R$
anomalies, and $H \times R$ anomalies can be computed at $\hat \phi_0$ using
$V$ instead of $T_{\hat \phi_0} \CM$. The result can be summarized by

\noindent {\bf Theorem 4:} Let $\CM \in V$ be a moduli space described by the
critical points of a flavor invariant superpotential $W$ of $R$-charge two. 
Then the anomalies of an unbroken subgroup $H \subseteq F \times R$ at a point
$\hat \phi_0 \in \CM$ can be computed using the entire space $V$, instead of
the tangent space of $\CM$. If the anomaly matching conditions between the UV
and IR theories for $H$ are  satisfied at $\hat \phi_0$, they are also
satisfied at all points of any moduli space $\CM' \in V$ given by the critical
points of any $W'$ (including $W'=0$ and $W'=W$), and at which $H$ is unbroken.

This result tells us that for moduli spaces described by a superpotential, the
precise form of the moduli space is irrelevant. The only role of possible
quantum deformations is to remove points of higher symmetry from the moduli
space.

One interesting application of this result is to prove that anomaly matching
conditions are compatible with integrating out heavy fields. Assume that one
has a theory with a moduli space $\mq$ described by a superpotential $W\left(
\hat \phi, \Lambda\right)$. Now perturb the UV theory by adding a tree level
mass term $U\left(\phi \right) = m_{ij} \phi^i \phi^j$ to the superpotential.
$U\left(\phi \right)$ is gauge invariant, and can be written as a polynomial
$W_m( \hat \phi )$ of the gauge invariant composites $\hat \phi$ of the IR
theory. If the UV theory contains no singlets, then $W_m(\hat \phi)$ is
linear in the basic gauge invariant composite fields $\hat \phi$. From this, it
immediately follows that the effective superpotential of the massive theory is
given by
\begin{equation}
W(\hat \phi,\Lambda) = W_0(\hat \phi,\Lambda) + W_m(\hat \phi),
\end{equation}
where $W_0$ is the superpotential in the absence of a mass term, since a linear
term in the fields is equivalent to a redefinition of the source.

The anomalies in the IR theory for any unbroken subgroup are unaffected by the
change in the moduli space due to the addition of the mass term. They are still
obtained by tracing over the whole space $V$. In the UV theory, one should
trace not over the whole space $U$, but only over the modes that remain
massless when $W_m$ is turned on. But it is easy to see that the massive modes
in the UV theory form a real representation of the unbroken symmetry. The
argument is the same as that used in the IR theory, except that $W_{ij}(\hat
\phi_0)$ is replaced by the (constant) matrix $m_{ij}$. Thus the mass term does
not introduce any modifications to the anomaly in the UV or IR theory. Thus one
finds that if the 't~Hooft conditions are verified for a theory, they are also
valid for any theory obtained by integrating out fields by adding a mass term.
The same argument is used to extend Theorem~3 to the case where there is a tree
level invariant superpotential $W$. Assume $H \subseteq F$ is unbroken at $\hat
\phi _0 \in \mc$, and $\phi_0 \in \pi^{-1}(\hat \phi _0) \subseteq U^W$
satisfies the conditions of theorem~3. Then theorem~4 applied to the $G
\times F$ invariant subset $U^W \subseteq U$ gives 
\begin{equation}
{\cal A}_{H^*}({T_{\phi_0}U^W}) = {\cal A}_{H^*}(U),
\end{equation}
because $\phi_0$ is a fixed point of the $H^*$ action  and $U^W$ is the set of
critical points of an $H^*$ invariant  superpotential. (Here ${\cal
A}_{H^*}(Z)$ denotes the $H^*$ anomaly computed  using the vector space $Z$.)
Thus
\begin{equation}
{\cal A}_H(T_{\hat \phi_0}\mc) = {\cal A}_{H^*}(T_{\phi_0}U^W) 
= {\cal A}_{H^*}(U) = {\cal A}_H(U),
\end{equation}
where the first and third equalities are shown in the proof of Theorem~3. 

We finally show how to use Theorem~4 to globalize the point-by-point  result of
the matching theorem for theories with a moduli space given by a
superpotential. For a generic supersymmetric gauge  theory, the representation
$\rho$ of the gauge group on $U$  is made up of $N_{F_i}$ tensor copies
(flavors) of representations  $\rho_i$, $i=1,\ldots,r$ and the non-anomalous
flavor symmetry group is 
\begin{equation}
F = SU(N_{F_1}) \times SU(N_{F_2}) \times \cdots \times 
SU(N_{F_s}) \times U(1)_1 \times U(1)_2 \times \cdots \times 
U(1)_{r-1} \times U(1)_R.  
\end{equation}
The most non-trivial anomaly check is to show that the anomalies for the entire
flavor group $F$ match at the origin. From this, anomaly matching at all points
of the moduli space follows by another application of Theorem~4. The
non-trivial anomalies are $SU(N_{F_i})^3$, $SU(N_{F_i})^2 U(1)_j$, and $U(1)_i
U(1)_j U(1)_k$, where the $U(1)_j$ include the $R$-symmetry. An $SU(N_{F_i})^3$
anomaly is proportional to the $d$-symbol for the $SU(N_{F_i})$ group. Since
the $d$-symbol is non-zero for $SU(3)$, the $SU(N_{F_i})$ anomaly can be
computed by looking at an $SU(3)$ subgroup. This is a standard trick for
computing the anomalies of representations of a general Lie group. To show that
the $SU(N_{F_i})^3$ anomaly matches, it is sufficient to find a point on the
moduli space that leaves an $SU(3)$ subgroup of $SU(N_{F_i})$ unbroken. The
$SU(N_{F_i})^3$ anomaly must then also match at the origin. Similarly, the
$SU(N_{F_i})^2 U(1)_j$ anomaly matching can be proven by considering a point on
the moduli space which leaves a $U(1)$ subgroup of $SU(N_{F_i})$ and $U(1)_j'$
unbroken. There is a subtlety here: the unbroken $U(1)_j'$ generator can be a
linear combination of the original $U(1)_j$ generator and some of the (broken)
$SU(N_{F_k})$ generators. However, having already proven that the
$SU(N_{F_i})^3$ anomaly matches,  $SU(N_{F_i})^2 U(1)_j'$ anomaly matching
implies $SU(N_{F_i})^2 U(1)_j$ anomaly matching. The $U(1)_i U(1)_j U(1)_k$
anomaly matching is proven by finding a point where three $U(1)'$'s are
unbroken, where again $U(1)'$ is a linear combination of the original $U(1)$'s
and $SU(N_{F_i})$ generators. This procedure might seem complicated, but it is
not that involved in practice. In the case of supersymmetric QCD, we will show
explicitly how one can prove anomaly matching for $N_F=N_c+1$ by considering
just one point on the moduli space (and its charge conjugate partner).

\section{Applications}\label{sec:examples}

The construction of the moduli space of supersymmetric gauge theories can be
highly non-trivial when there are many gauge invariant combinations of the
fundamental fields. The results of the preceding sections help simplify the
analysis of the structure of the moduli space, and of anomaly matching between
the fundamental and massless composite fields. We treat the following problems:
\begin{itemize}
\item[(A)] Determining the flavor  isotropy group $F_{\pi(\phi_0)}$ of the
vacuum $\pi(\phi _0) \in \mc$, i.e, the maximal unbroken subgroup of $F$ at
$\pi(\phi_0)$.\footnote{This  problem is not trivial when the basic gauge
invariants of the theory are not known.}
\item[(B)] Setting sufficient conditions for the  $F_{\pi(\phi_0)}$ anomalies
in $U$ to match  the corresponding anomalies in $T_{\pi(\phi_0)} \mc$
\item[(C)] Setting  sufficient conditions for  anomalies to match at {\em
every} vacuum of the (possibly  quantum deformed) moduli space.
\end{itemize}
Three different situations can be considered:
\begin{itemize}
\item[(i)] The basic gauge  invariants of the given theory are not known. 
\item[(ii)] The basic gauge invariants are known, but not the  constraints
among them.
\item[(iii)] Both the invariant and the constraints  are known. 
\end{itemize}

In situation~(i) theorems~2 and~3 tell us that if $\phi_0 \in U^W$ is 
smooth, totally breaks the complexification $G$ of the gauge group $G_r$,  and
has a closed orbit $G \phi_0$ (equivalently, there is a D-flat  point in $G
\phi_0$), then the anomalies do match between $U$  and $T_{\pi(\phi_0)}\mc$, 
$$ A_{F_{\hat \phi _0}}(T_{\hat \phi _0} \mc) =  A_{F_{\hat \phi _0}}(U).$$ 
This solves problem~B. To determine the  flavor isotropy group
$F_{\pi(\phi_0)}$ we note from  Theorem~2 in Section~\ref{sec:vacua}
 that for $\phi_0$ as
above $T_{\phi_0}(\pi^{-1}(\pi(\phi_0))) = \lie G \phi_0$,  and
$F_{\pi(\phi_0)}$ is isomorphic to the  $G \times F$ isotropy group at
$\phi_0$.\footnote{Note  that in general $F_{\phi_0} \subseteq
F_{\pi(\phi_0)}$.}

In situation~(ii) we can solve~A and~B in the same way  as in situation~(i),
but we have the alternative of directly determining $F_{\pi(\phi_0)}$, for
which only the gauge invariants  and not the constraints among them are needed.

In situation (iii) we have an additional way of dealing with  problem~B, using
the  fact that the gauge group $G_r$ is semisimple. As $\mc$ is known  we can
calculate its dimension. Suppose dim $\mc = $ dim $U - d_G$  and a point
$\phi_0$ is found such that: $\pi(\phi _0)$ is  smooth,  $\phi_0 $  breaks
completely $G$. Then the $F_{\pi(\phi _0)}$  anomalies match between $U$ and
$T_{\pi(\phi _0)} \mc$. This follows from theorems~1, 2 and 3. Note that it
is crucial  that $\phi_0$ break completely the {\em complexification} of the
gauge  group. Consider the example of supersymmetric QCD with $N_F = N_c -1$.
The classical moduli space is the span of the unconstrained fields  $M^i_j$, so
$\mc \cong \c ^{N_F^2}$ and dim $\mc = $ dim $U - d_G$.  The point $\phi_0$ of
coordinates $(\tilde{Q}_0)_{\alpha i} = 0 , \;  Q_0^{i \alpha}~=~m~\delta^{i
\alpha}, \alpha \leq N_F, 0$ for $\alpha =  N_c$, totally breaks $G_r = 
SU(N_c)$, but not $G = SL(N_c,\c)$. $\mc$ is smooth everywhere,  in particular 
$\pi(\phi_0) = 0$ is a smooth point. However, anomalies do not match  between
the UV and $T_{0} \mc = \text{span}(M^i_j)$. Now consider  the point $\phi_1$
with coordinates $(\tilde{Q}_1)_{\alpha i} =  Q_1^{i \alpha} = m \delta^{i
\alpha},  \alpha \leq N_F, 0$ for $\alpha = N_c$. This point breaks completely 
the {\em complexification} of the gauge group, and it is mapped onto a  smooth
point. Theorems~1, 2 and 3 predict anomaly matching,  which is
straightforward from the fact that $F_{\pi(\phi_1)}$  is the diagonal
$SU(N_F)$. \\ Theorem~4 in  Section~\ref{sec:ir} 
 gives an answer to problem C. Whenever
$\mc$ is the set  of critical points of an invariant superpotential,  the
matching of the $F_{\hat \phi _i}$  anomalies at the vacua $\hat \phi _i,
i=1,...,s$ implies the matching of  $F_{\hat \phi _i}$ anomalies at any vacuum
$\hat \phi$  where $F_{\hat \phi} \supseteq F_{\hat \phi _i}$.  In particular,
we can take $\hat \phi = 0 \in \mc$, where the flavor  group is not broken at
all. The points $\hat \phi _i$  can be chosen such that  the matching of the
$F_{\hat \phi _i}$ anomalies at the origin  (where the tangent space is the
full ambient vector  space) implies the matching of the full flavor group 
anomalies there. Applying again Theorem~4 we prove that every point of $\mc$
will pass 't~Hooft's test. This extends to $\mq$ if $\mq$ also comes from  a
flavor invariant superpotential. This idea can be used to prove anomaly 
matching for the large family of s-confining theories  of ref~\cite{sconf}, and
for those obtained from  them by integrating out any number of flavors. The
latter may have  a quantum modified moduli space, which can be  embedded in the
same ambient vector space of the s-confining  theory, and is described with an
invariant superpotential (Section~\ref{sec:ir}).\\ Below, we extend the  
explicit anomaly
matching calculations in supersymmetric  QCD done in \cite{seiberg} to  a
generic point in moduli space, and show that anomalies  match precisely at
those points predicted by our theorems.  The amount of calculations using both
approaches is contrasted. 

\subsection{Supersymmetric QCD} 

\subsubsection{$N_c=2$}

The quarks and antiquarks transform as the ${\bf 2}$ of $SU(2)$, and so can be
treated together as $2N_F$ flavors of doublets. The flavor symmetry is
$SU(2N_F) \times U(1)_R$. The basic gauge invariant field is the meson
\begin{equation}
V^{ij} = \epsilon_{\alpha \beta} Q^{i \alpha} Q^{j \beta}.
\end{equation}
The transformation properties of the fundamental fields and composites under
the flavor symmetry is listed in Table~\ref{table:n=2}.

\begin{table}[tbph]
\caption{Flavor representation of the fundamental and composite fields for 
$SU(2)$ gauge theory with $2N_F$ fundamentals.}
\begin{tabular}{ccc}
 & $SU(2N_F)$ & $U(1)_R$ \\
\tableline
$Q^{i\alpha}$ & $\sqr$ & ${N_F-2 \over N_F}$ \\
$V^{ij}$ & $\aa$ & $2 {N_F-2 \over N_F}$ \\
\end{tabular}
\label{table:n=2}
\end{table}

The classical moduli space for $N_F=1$ (i.e. $N_F < N_C$) is given by all
possible values for $V^{ij}$. Since $V^{ij}$ is a $2 \times 2$ antisymmetric 
matrix, it has the form
\begin{equation}\label{exam:1}
V^{ij} = \left(\begin{array}{cc}
0 & v \\
-v & 0 \\
\end{array}\right).
\end{equation}
For $v\not =0$, the unbroken flavor symmetry is flavor $SU(2)$, whereas for
$v=0$, the full $SU(2) \times U(1)_R$ symmetry is unbroken. It is
straightforward to verify that the $SU(2)^3$ and $SU(2)$ anomalies match
between the UV and IR theories so that the IR and UV theories have the same
anomalies at a point on the moduli space where $v\not=0$. At the origin of
moduli space $v=0$, the UV and IR anomalies do not match. The anomaly matching
theorem can be used at all points $v\not=0$. One can pick a point
\begin{equation} 
Q^{i \alpha}_0 = w 
\left(\begin{array}{cc} 1 & 0 \\ 
0 & 1 \\
\end{array}\right) 
\end{equation}
with $v=w^2$, which gives Eq.~(\ref{exam:1}) for $V^{ij}$. The orbit containing
$Q_0^{i \alpha}$ is closed, and $G$ is completely broken, so the UV and IR
anomalies must match. There is no point above the origin $V^{ij}=0$ that
satisfies the requirements for applying the anomaly matching theorem. Thus for
$N_F=1$, all points where the anomalies match can be described using the
theorem. Note that the point
\begin{equation} 
Q^{i \alpha}_0 = w 
\left(\begin{array}{cc} 1 & 0 \\ 
0 & 0 \\
\end{array}\right) 
\end{equation}
completely breaks the gauge group $G_r$, but not the complexified gauge group
$G$. Completely breaking $G_r$ is not sufficient to use theorem~3.

For $N_F \ge 2$, the classical moduli space is the set of all $V^{ij}$'s
subject to the constraint that ${\rm rank}\, V \le 2$~\cite{seiberg}.  At a
generic point
\begin{equation}\label{exam:2}
V^{ij} = \left(\begin{array}{cc|ccc}
0 & v & 0 & \cdots & 0\\
-v & 0 & 0 & \cdots & 0\\
\hline
0 & 0 & 0 & \cdots & 0\\
\vdots & \vdots & \vdots & & \vdots\\
0 & 0 & 0 & \cdots & 0\\
\end{array}\right),
\end{equation}
with $v\not=0$, the flavor group is broken to $SU(2) \times SU(2N_F-2) \times
U(1)_R$. To compute the anomalies in the UV and IR theories, it is convenient
to break up the flavor index into $i=1,2$ and $i=3,\ldots 2N_F$. The quarks can
be broken into $Q_1$ and $Q_2$, respectively. The meson $V^{ij}$ can be written
as
\begin{equation}\label{exam:3}
V^{ij} = \left(\begin{array}{cc }
V_{11} & V_{12} \\
-V_{12}^T & V_{22}\\
\end{array}\right),
\end{equation}
where $V_{11}$ and $V_{22}$ are $2\times 2$ and $(2N_F-2) \times (2N_F-2)$
antisymmetric matrices, and $V_{12}$ is a $2 \times 2N_F-2 $ matrix. Denoting
the tangent vectors to the moduli space by $\delta V$, one sees that the
constraint ${\rm rank}\, V \le 2$ requires that $\delta V_{22}=0$ if $v\not=0$.
The flavor transformations of the fields is given in Table~\ref{table:n=2f}.

\begin{table}[tbph]
\caption{Flavor representation of the fundamental and composite fields for 
$SU(2)$ gauge theory with $2N_F$ fundamentals under the 
$SU(2) \times SU(2N_F-2) \times U(1)_R$ subgroup.}
\begin{tabular}{cccc}
 & $SU(2)$ & $SU(2N_F)$ & $U(1)_R$ \\
\tableline
$Q^{i\alpha}_1$ & $\sqr$ & $-$ & $0$ \\
$Q^{i\alpha}_2$ & $-$ & $\sqr$ & ${N_F-2 \over N_F -1}$ \\
$\delta V^{ij}_{11}$ & $-$ & $-$ & $0$ \\
$\delta V^{ij}_{12}$ & $\sqr$ & $\sqr$ & ${N_F-2 \over N_F-1}$ \\
\end{tabular}
\label{table:n=2f}
\end{table}

It is easy to verify that the flavor anomalies of $Q_1$ and $Q_2$ are the same
as those of $\delta V_{11}$ and $\delta V_{12}$,  so that the flavor anomalies
match at all points where $v\not=0$~\cite{seiberg}. At the origin $V^{ij}=0$,
the full flavor group is unbroken, and the tangent vectors $\delta V^{ij}$ are
unconstrained. The anomalies in the UV and IR theories are computed using the
fields in Table~\ref{table:n=2}, and match only for the case of
$N_F=3$~\cite{seiberg}.

The point
\begin{equation}
Q_0 = w\left( \begin{array}{ccccc}
1 & 0 & 0 & \cdots & 0\\
0 & 1 & 0 & \cdots & 0\\
\end{array}\right)
\end{equation}
with $w^2=v$ projects to a point Eq.~(\ref{exam:2}) on the classical moduli
space. As before, $Q_0$ is a point on a closed orbit that completely breaks the
gauge group, so anomaly matching is guaranteed. This explains all the points
where the anomalies match for $N_F\not=3$. In the case where $N_F=3$, the
classical moduli space is described by a superpotential \begin{equation} W
\propto {\rm Pf}\, V \end{equation} In this case, Theorem~4 implies that
$SU(2) \times SU(2N_F-2) \times U(1)_R$ anomaly matching at $V^{ij}\not =0$
also implies that the anomalies for this subgroup match at the origin. But that
is sufficient to guarantee anomaly matching for the full $SU(2N_F) \times
U(1)_R$ flavor symmetry at the origin.

\subsubsection{$N_c>2$}

The analysis can be repeated for the case $N_c>2$. The computations are more 
involved than for $N_c=2$, because it is tedious to find the tangent vectors at
a given point on the moduli space. One can show that all points where anomalies
match between the UV and IR theories are covered by the anomaly matching
theorem, except for trivial cases in which the unbroken flavor symmetry group
is anomaly-free. Instead of going through a detailed description of anomaly
matching at the different points of the moduli space, we will illustrate the
anomaly matching at one interesting point (the ``baryon point'') $\hat \phi_0$
on the moduli space for $N_F \ge N_c$, where $M^i_j=0$, $B^{12\cdots N_c}=1$
(with all other components zero), $\tilde B=0$.

The unbroken flavor group at $\hat \phi_0$ is $SU(N_c)_L \times SU(N_F-N_c)_L
\times SU(N_F)_R \times U(1)_B \times U(1)_R$. The point $\phi_0$
\begin{equation}
Q_0 = \left( \begin{array}{cccc|ccc}
1 & 0 & 0 & 0 & 0 & \cdots & 0\\
0 & 1 & 0 & 0 & 0 & \cdots & 0\\
\vdots & & & & & & \vdots\\
0 & 0 & 0 & 1 & 0 & \cdots & 0 \\
\end{array}\right),\qquad \tilde Q_0 =0,
\end{equation}
projects to $\hat \phi_0$, $\pi(\phi_0) = \hat \phi_0$. Theorems~2 and~3 tells
us that the anomalies should match at this point. The tangent vectors on the
classical moduli space at $\hat \phi_0$ can be found easily in this case, since
$\pi'_\phi$ is onto. They are given by $\delta M^i_j$, $i \le N$, $\delta
B^{1\ldots N_c }$ and $\delta B^{1\ldots \hat k \ldots N_c r}$, where $1 \le k
\le N_c$, $r > N_c$, and $\hat k$ means that the value $k$ is omitted. The
transformation properties of the fundamental and composite fields is given in
Table~\ref{table:nc}. The anomalies in the UV and IR theories are tabulated in
Table~\ref{table:anom}. Clearly, it is simpler to use the anomaly matching
theorem, instead of computing the entries in Table~\ref{table:anom}.

\begin{table}[tbph]
\caption{Flavor transformations of the fundamental and composite fields at a
``baryon point.'' The quark fields are divided up into $Q_1^{i \alpha}$ and
$Q_2^{i \alpha}$, which are $Q^{i\alpha},\ i\le N_c$ and $Q^{i \alpha}\
i>N_c$, respectively. The tangent vectors on the moduli space are $\delta
M^i_j,\ i \le N_c$, $\delta B^{1\ldots N }$ and $\delta B^{1\ldots \hat k
\ldots N_c r},\ r>N_c,\ k \le N_c$. }
\begin{tabular}{cccccc}
 & $SU(N_c)_L$ & $SU(N_F-N_c)_L$ & $SU(N_F)_R$ & $B$ & $R$ \\
\tableline
$Q_1^{i\alpha}$& $\sqr$ & $-$ & $-$ & $0$ & $0$ \\
$Q_2^{i\alpha}$ & $-$ & $\sqr$ & $-$ & $- N_F$ & 
$(2N_F-2N_c)/(2N_F-N_c)$ \\
$\tilde Q_{j \alpha}$ & $-$ & $-$ & $\overline{\sqr}$ & $N_F-N_c$ & 
$(2N_F-2N_c)/(2N_F-N_c)$ \\
\tableline
$\delta M^i_j$ & $\sqr$ & $-$ & $\overline{\sqr}$ & $N_F-N_c$ & 
$(2N_F-2N_c)/(2N_F-N_c)$\\
$\delta B^{1\ldots N }$ & $-$ & $-$ & $-$ & $0$ & $0$ \\
$\delta B^{1\ldots \hat k \ldots N_c r} $ & $\overline{\sqr}$ & $\sqr$ & $-$ &
$-N_F$ & $(2N_F-2N_c)/(2N_F-N_c)$ \\
\end{tabular}
\label{table:nc}
\end{table}

\begin{table}[tbph]
\def\nfc{s}
\def\ri{t} 
\caption{Flavor anomalies in the UV and IR theories due to the various fields.
$\lambda$ is the gaugino. Here $s=N_F-N_c$, $t=-N_c/(2N_F-N_c)$. See the
Table~\protect{\ref{table:nc}} caption for the definition of the various
fields.}
\begin{tabular}{c|cccc|ccc}
&\multicolumn{4}{c|}{UV} & \multicolumn{3}{c}{IR} \\
\cline{1-8}
& $Q_1^{i\alpha}$ & $Q_2^{i \alpha}$ & $\tilde Q_{j \alpha}$ &
$\lambda$ & $\delta M^i_j$ & $\delta B^{1\ldots N }$ & 
$\delta B^{1\ldots \hat k \ldots N_c r}$ \\
\tableline
$SU(N_c)^3_L$ & $N_c$ & 0 & 0 & 0 & $N_F$ & 0 & $-\nfc$ \\
$SU(\nfc)^3_L$ & 0 & $N_c$ & 0 & 0 & 0 & 0 & $N_c$ \\
$SU(N_F)^3_R$ & 0 & 0 & $-N_c$ & 0 & $-N_c$ & 0 & 0 \\
$B^3$ & 0 & $-N_c\nfc N_F^3$ & $N_c N_F \nfc^3$ & 0 &
$N_c N_F\nfc^3$ & 0 & $-N_c\nfc N_F^3$ \\
$R^3$ & $-N_c^2$ & $N_c \nfc \ri^3$ & $N_c N_F \ri^3$ & $N_c^2-1$ & 
$N_c N_F \ri^3$ & $-1$ & $N_c \nfc \ri^3$ \\
$SU(N_c)^2_L\ B$ & 0 & 0 & 0 & 0 & $N_F \nfc$ & 0 & $-\nfc N_F$ \\ 
$SU(\nfc)^2_L\ B$ & 0 & $-N_c N_F$ & 0 & 0 & 0 & 0 & $- N_c N_F$ \\
$SU(N_F)^2_R\ B$ & 0 & 0 & $N_c \nfc$ & 0 & $N_c \nfc$ & 0 & 0\\
$SU(N_c)^2_L\ R$ & $-N_c$ & 0 & 0 & 0 & $N_F \ri$ & 0 & $\nfc \ri$ \\ 
$SU(\nfc)^2_L\ R$ & 0 & $N_c \ri$ & 0 & 0 & 0 & 0 & $N_c \ri$ \\
$SU(N_F)^3_R\ R$ & 0 & 0 & $N_c \ri$ & 0 & $N_c \ri$ & 0 & 0 \\
$B^2\ R$ & 0 & $N_c \nfc N_F^2 \ri$ & $N_c N_F \nfc^2 \ri$ & 0 & $N_c N_F
\nfc^2 \ri$ & 0 & $N_c \nfc N_F^2 \ri$ \\
$R^2\ B$ & 0 & $-N_c N_F \nfc \ri^2$ & $N_c N_F \nfc \ri^2$ & 0 & 
$N_c N_F \nfc \ri^2$ & 0 & $-N_c \nfc N_F \ri^2$ \\
$B$ & 0 & $- N_c \nfc N_F$ & $N_c N_F \nfc$ & 0 & $N_c N_F \nfc$ & 0 & $-N_c
\nfc N_F$ \\
$R$ & $-N_c^2$ & $N_c \nfc \ri$ & $N_c N_F \ri$ & $N_c^2 -1$ & 
$N_c N_F \ri$ & $-1$ & $N_c \nfc \ri$ \\
\end{tabular}
\label{table:anom}
\end{table}

\subsubsection{S-confinement and Quantum Deformations}

For any value of $N_c$ and $N_F = N_c + 1$ the classical moduli space  of
supersymmetric QCD  is the set of critical points of the flavor invariant
superpotential  $M^i_j B_i \tilde{B}^j - \text{det} M$. The effective
superpotential  is therefore~\cite{susy} 
\begin{equation}
W_0 = \frac{1}{\Lambda ^{2N_c -1}} \left( 
M^i_j B_i \tilde{B}^j - \text{det} M \right), 
\end{equation}
and the quantum moduli space $\mq$ agrees with $\mc$.  The matching of the
anomalies of the  $SU(N_c)_L \times SU(N_F - N_c)_L \times SU(N_F)_R  \times
U(1)_B \times U(1)_R$ flavor subgroup unbroken  at the ``baryon'' point of the
previous subsection, together  with that of the ``antibaryon'' point (with $Q
\leftrightarrow \tilde{Q}$)  with unbroken flavor subgroup  $SU(N_c)_R \times
SU(N_F - N_c)_R \times SU(N_F)_L  \times U(1)_B \times U(1)_R$ imply  the
matching for the full flavor group  at the origin (Theorem~4). Another
application of Theorem~4 proves 't~Hooft's conditions are satisfied at  any
vacuum in the moduli space. According to the discussion of  
Section~\ref{sec:ir}, adding
a tree level mass term $m Q^{N_c+1 \alpha} \tilde{Q}  _{\alpha N_c+1}$ to $W$
gives the effective superpotential
\begin{equation} \label{nf=na}
W_m = \frac{1}{\Lambda ^{2N_c -1}} \left( 
M^i_j B_i \tilde{B}^j - \text{det} M \right) + m M^{N_c+1}_{N_c+1}.
\end{equation}
The critical points of Eq.~(\ref{nf=na}) give the quantum moduli  space
\begin{eqnarray}
M^i_{N_c+1} &=& M^{N_c +1}_j = 0, \nonumber \\
 B_i &=& \tilde{B}^i = 0, \; i \leq N_c , \nonumber \\
\text{det} M - B \tilde{B} &=& \Lambda_N^{2N +1},
\end{eqnarray}
of the $N_F = N_c$ theory, where $B = B_{N_c + 1}, \tilde{B} =  \tilde{B}^{N_c
+ 1}$ and the determinant extends to  the light flavors $i,j \leq N_c$.
Applying Theorem~4 to $W_0$ and $W_m$ proves that anomaly matching  for the
$N_F = N_c + 1$ theory {\em implies} anomaly matching  {\em at every point of
the quantum deformed  moduli space of the $N_F = N_c$ theory}. We can therefore
avoid the  explicit checks done at isolated points of $\mq$ in \cite{seiberg} 
and be sure the result holds at every point. Note that,  although $\mq$ of the
$N_F = N_c$ theory can be described  as the set of critical points of a
superpotential when  embedded in the vector space of gauge invariants of the
$N_F = N_c +1$  theory, $\mc$ cannot. In particular, anomalies do not match at 
the point $M^i_j = 0, \; B = \tilde{B} = 0$, which belongs  to $\mc$ but not to
$\mq$. 

\subsection{Other S-confining Theories}

Our analysis for QCD extends to other s-confining theories,  an exhaustive list
of which can be found in \cite{sconf}.  For all s-confining theories $\mc =
\mq$ can be  described by a flavor invariant superpotential, so a  finite
number of points $\phi_i \in U$  satisfying the hypothesis of the matching 
theorem is enough to prove anomaly matching at {\em every}  point of their
moduli space and at the quantum deformed moduli  space of the theories obtained
from them by integrating out  a heavy fields. Explicit verification of this
matching both  for the s-confining theory and the quantum deformed one  is a
formidable task. Even the determination of a complete  set of basic gauge
invariants is sometimes nontrivial.  We give one simple example of an
s-confining theory.

\subsubsection{$SP(2N,\c)$ with $2N + 4$ fundamentals $Q^{\alpha i}$}

$SP(2N,\c)$ is the group that leaves invariant the 2-form
\begin{equation}
J_{\alpha \beta} = \left( \begin{array}{cc} 
0 & {\cal I}_{{N \times N}} \\
-{\cal I}_{{N \times N}} & 0 \end{array} \right). 
\end{equation} 
This is the complexification of the group $G_r = SP(2N) = SP(2N,\c)  \cap
U(2N)$. Just one point $\phi_0$ is required to prove  anomaly matching in the
s-confining $SP$ theory with  $2N + 4$ fundamentals $Q^{\alpha i}$, 
$Q_0^{\alpha i} = m \delta^{\alpha i}, i \leq 2N, 0$ for  $i > 2N$. This point
totally breaks $G$, and its  orbit is the closed set $SP(2N,\c)$, naturally 
embedded in the vector space $\c ^{2N(2N+4)}$ of $2N \times (2n+4)$  matrices. 

\section{Conclusions}

For supersymmetric gauge theories with a reductive gauge group  $G$, the
classical  moduli space $\mc$ is the algebraic quotient of the  algebraic set
of critical points of the superpotential  under the action of $G$. There are
known bounds  for its dimension, which can be determined without  knowing the
basic invariant polynomials when  closed orbits of maximum dimension are
found.  The anomaly matching theorem can be used to show  that the 't~Hooft
consistency conditions are satisfied at points that totally break the
complexification  of the gauge group and either have a closed orbit or are
mapped onto a smooth  point of $\mc$. Anomalies will match at {\em every} point
of $\mq$ if a  few suitable points satisfying the above hypothesis are found
and  both $\mc$ and $\mq$ can be described as the set of critical points  of a
flavor invariant superpotential.  Anomaly matching for a theory {\em implies}
anomaly  matching for those theories obtained from it by integrating  out
matter, even when these have a quantum modified  moduli space. Anomalies match 
for the large family of s-confining theories  and those obtained from them by
integrating out a flavor, which  have a quantum modified moduli space. The
explicit anomaly computations  found in the literature can often be avoided, if
one uses the results discussed here. These results also allow one to anticipate
if $\mc$ or a  quantum deformed $\mq$ describes correctly the IR massless
modes,  or if an alternative description (such us a dual theory)  is required. 
Extensions of our results to  cases where the complexification of the gauge
group is not  totally broken and applications to dual theories is currently
under study.

\section{Acknowledgments}

We are indebted to M.L.~Barberis, M.~Hunziker and N.~Wallach,for extensive
discussions on algebraic geometry, and to N.~Wallach for giving us a copy of
his book~\cite{gw} prior to publication. We are also grateful to G.~Schwarz for
telling us about Knop's theorem, and for discussion about when $\pi'$ can be
surjective. This work was supported in part by a Department of Energy grant
DOE-FG03-97ER40546.

\end{document}